\documentclass[12pt]{article}
\usepackage{setspace}
\usepackage{amssymb}
\usepackage{amsmath,amsfonts}
\usepackage{graphicx}
\usepackage{fullpage}
\usepackage{color}
\usepackage{hyperref}
\newcommand{\nn}{\nonumber\\}

\newcommand{\be}{\begin{equation}}
\newcommand{\ee}{\end{equation}}
\newcommand{\bea}{\begin{eqnarray}}
\newcommand{\eea}{\end{eqnarray}}

\newcommand{\eq}{&=&}
\newcommand{\commute}[2]{\left[ #1 \, , \, #2 \right]}
\newcommand{\anticommute}[2]{\left\{ #1 \, , \, #2 \right\}}
\newcommand{\ket}[1]{\left\lvert #1 \right\rangle}
\newcommand{\mfQ}{\mathfrak{Q}}
\newcommand{\mfQd}{\mathfrak{Q}^\dagger}
\newcommand{\mfS}{\mathfrak{S}}
\newcommand{\mft}{\mathfrak{t}}
\newcommand{\mfSd}{\mathfrak{S}^\dagger}
\newcommand{\mcJ}{\mathcal{J}}
\newcommand{\half}{\frac{1}{2}}
\newcommand{\tensor}[4]{\left\lvert #1 \right\rangle_F\times\left\lvert #2 \right\rangle_B\times \left\lvert\psi_{ #4 }^{ #3 }\right\rangle}
\newcommand{\upa}{\uparrow}
\newcommand{\dowa}{\downarrow}
   
\newcommand{\marcshapirowbox}
{\stln \lower2.6ex\hbox{
\begin{picture}(12.6,2.6)
\multiput(.3,1.3)(1,0){3}{\sfr}
\put(3.3,1.3){\framebox(3,1){$\cdots$}}
\put(6.3,1.3){\sfr}
\put(7.3,1.3){\sfr}
\put(8.3,1.3){\framebox(3,1){$\cdots$}}
\put(11.3,1.3){\sfr}
\put(7.3,1){$\underbrace{~~~~~~~~~~~~~~~}_{2s}$}
\multiput(.3,.3)(1,0){3}{\sfr}
\put(3.3,.3){\framebox(3,1){$\cdots$}}
\put(6.3,.3){\sfr}
\put(.3,0){$\underbrace{~~~~~~~~~~~~~~~~~~~~~}_{n}$}
\end{picture}}}

\newcommand{\marcshapirowboxs}
{\stln \lower2.6ex\hbox{
\begin{picture}(7.6,2.6)
\multiput(.3,1.3)(1,0){3}{\sfr}
\put(3.3,1.3){\framebox(3,1){$\cdots$}}
\put(6.3,1.3){\sfr}
\put(.3,.3){\sfr}
\end{picture}}}

\usepackage{multirow}
\usepackage[mathscr]{eucal}
\usepackage{amsfonts}
\usepackage[mathscr]{eucal}
\newcommand{\stln}{\setlength{\unitlength}{2.2ex}}
\newcommand{\fr}{\framebox(1,1){}}
\newcommand{\sfr}{\framebox(1,1){\begin{picture}(1,1)
  \put(0,0){\line(1,1){1}}\end{picture}}}

\newcommand{\onebox}
{\stln \lower1.4ex\hbox{
\begin{picture}(1.6,1.6)
\put(.3,.3){\fr}
\end{picture}}}

\newcommand{\twobox}
{\stln \lower1.4ex\hbox{
\begin{picture}(2.6,1.6)
\put(.3,.3){\fr}
\put(1.3,.3){\fr}
\end{picture}}}

\newcommand{\threebox}
{\stln \lower1.4ex\hbox{
\begin{picture}(3.6,1.6)
\multiput(.3,.3)(1,0){3}{\fr}
\end{picture}}}

\newcommand{\fourbox}
{\stln \lower1.4ex\hbox{
\begin{picture}(4.6,1.6)
\multiput(.3,.3)(1,0){4}{\fr}
\end{picture}}}

\newcommand{\fivebox}
{\stln \lower1.4ex\hbox{
\begin{picture}(5.6,1.6)
\multiput(.3,.3)(1,0){5}{\fr}
\end{picture}}}

\newcommand{\sixbox}
{\stln \lower1.4ex\hbox{
\begin{picture}(6.6,1.6)
\multiput(.3,.3)(1,0){6}{\fr}
\end{picture}}}

\newcommand{\genrowbox}
{\stln \lower1.4ex\hbox{
\begin{picture}(7.6,1.6)
\multiput(.3,.3)(1,0){3}{\fr}
\put(3.3,.3){\framebox(3,1){$\cdots$}}
\put(6.3,.3){\fr}
\end{picture}}}

\newcommand{\oneonebox}
{\stln \lower2.6ex\hbox{
\begin{picture}(1.6,2.6)
\put(.3,.3){\fr}
\put(.3,1.3){\fr}
\end{picture}}}

\newcommand{\twoonebox}
{\stln \lower2.6ex\hbox{
\begin{picture}(2.6,2.6)
\put(.3,1.3){\fr}
\put(1.3,1.3){\fr}
\put(0.3,0.3){\fr}
\end{picture}}}

\newcommand{\threeonebox}
{\stln \lower2.6ex\hbox{
\begin{picture}(3.6,2.6)
\multiput(.3,1.3)(1,0){3}{\fr}
\put(.3,.3){\fr}
\end{picture}}}

\newcommand{\fouronebox}
{\stln \lower2.6ex\hbox{
\begin{picture}(4.6,2.6)
\multiput(.3,1.3)(1,0){4}{\fr}
\put(.3,.3){\fr}
\end{picture}}}

\newcommand{\fiveonebox}
{\stln \lower2.6ex\hbox{
\begin{picture}(5.6,2.6)
\multiput(.3,1.3)(1,0){5}{\fr}
\put(.3,.3){\fr}
\end{picture}}}

\newcommand{\sixonebox}
{\stln \lower2.6ex\hbox{
\begin{picture}(6.6,2.6)
\multiput(.3,1.3)(1,0){6}{\fr}
\put(.3,.3){\fr}
\end{picture}}}

\newcommand{\twotwobox}
{\stln \lower2.6ex\hbox{
\begin{picture}(2.6,2.6)
\put(.3,.3){\fr}
\put(.3,1.3){\fr}
\put(1.3,.3){\fr}
\put(1.3,1.3){\fr}
\end{picture}}}

\newcommand{\threetwobox}
{\stln \lower2.6ex\hbox{
\begin{picture}(3.6,2.6)
\multiput(.3,1.3)(1,0){3}{\fr}
\put(.3,.3){\fr}
\put(1.3,.3){\fr}
\end{picture}}}

\newcommand{\fourtwobox}
{\stln \lower2.6ex\hbox{
\begin{picture}(4.6,2.6)
\multiput(.3,1.3)(1,0){4}{\fr}
\put(.3,.3){\fr}
\put(1.3,.3){\fr}
\end{picture}}}

\newcommand{\fivetwobox}
{\stln \lower2.6ex\hbox{
\begin{picture}(5.6,2.6)
\multiput(.3,1.3)(1,0){5}{\fr}
\put(.3,.3){\fr}
\put(1.3,.3){\fr}
\end{picture}}}

\newcommand{\sixtwobox}
{\stln \lower2.6ex\hbox{
\begin{picture}(6.6,2.6)
\multiput(.3,1.3)(1,0){6}{\fr}
\put(.3,.3){\fr}
\put(1.3,.3){\fr}
\end{picture}}}

\newcommand{\threethreebox}
{\stln \lower2.6ex\hbox{
\begin{picture}(3.6,2.6)
\multiput(.3,1.3)(1,0){3}{\fr}
\multiput(.3,.3)(1,0){3}{\fr}
\end{picture}}}

\newcommand{\fourthreebox}
{\stln \lower2.6ex\hbox{
\begin{picture}(4.6,2.6)
\multiput(.3,1.3)(1,0){4}{\fr}
\multiput(.3,.3)(1,0){3}{\fr}
\end{picture}}}

\newcommand{\fivethreebox}
{\stln \lower2.6ex\hbox{
\begin{picture}(5.6,2.6)
\multiput(.3,1.3)(1,0){5}{\fr}
\multiput(.3,.3)(1,0){3}{\fr}
\end{picture}}}

\newcommand{\sixthreebox}
{\stln \lower2.6ex\hbox{
\begin{picture}(6.6,2.6)
\multiput(.3,1.3)(1,0){6}{\fr}
\multiput(.3,.3)(1,0){3}{\fr}
\end{picture}}}

\newcommand{\fourfourbox}
{\stln \lower2.6ex\hbox{
\begin{picture}(4.6,2.6)
\multiput(.3,1.3)(1,0){4}{\fr}
\multiput(.3,.3)(1,0){4}{\fr}
\end{picture}}}

\newcommand{\fivefourbox}
{\stln \lower2.6ex\hbox{
\begin{picture}(5.6,2.6)
\multiput(.3,1.3)(1,0){5}{\fr}
\multiput(.3,.3)(1,0){4}{\fr}
\end{picture}}}

\newcommand{\sixfourbox}
{\stln \lower2.6ex\hbox{
\begin{picture}(6.6,2.6)
\multiput(.3,1.3)(1,0){6}{\fr}
\multiput(.3,.3)(1,0){4}{\fr}
\end{picture}}}

\newcommand{\fivefivebox}
{\stln \lower2.6ex\hbox{
\begin{picture}(5.6,2.6)
\multiput(.3,1.3)(1,0){5}{\fr}
\multiput(.3,.3)(1,0){5}{\fr}
\end{picture}}}

\newcommand{\sixfivebox}
{\stln \lower2.6ex\hbox{
\begin{picture}(6.6,2.6)
\multiput(.3,1.3)(1,0){6}{\fr}
\multiput(.3,.3)(1,0){5}{\fr}
\end{picture}}}

\newcommand{\sixsixbox}
{\stln \lower2.6ex\hbox{
\begin{picture}(6.6,2.6)
\multiput(.3,1.3)(1,0){6}{\fr}
\multiput(.3,.3)(1,0){6}{\fr}
\end{picture}}}

\newcommand{\oneoneonebox}
{\stln \lower3.8ex\hbox{
\begin{picture}(1.6,3.6)
\multiput(.3,.3)(0,1){3}{\fr}
\end{picture}}}

\newcommand{\twooneonebox}
{\stln \lower3.8ex\hbox{
\begin{picture}(2.6,3.6)
\multiput(.3,.3)(0,1){3}{\fr}
\put(1.3,2.3){\fr}
\end{picture}}}

\newcommand{\twotwoonebox}
{\stln \lower3.8ex\hbox{
\begin{picture}(2.6,3.6)
\multiput(.3,.3)(0,1){3}{\fr}
\put(1.3,1.3){\fr}
\put(1.3,2.3){\fr}
\end{picture}}}

\newcommand{\twotwotwobox}
{\stln \lower3.8ex\hbox{
\begin{picture}(2.6,3.6)
\multiput(.3,.3)(0,1){3}{\fr}
\multiput(1.3,.3)(0,1){3}{\fr}
\end{picture}}}

\newcommand{\threeoneonebox}
{\stln \lower3.8ex\hbox{
\begin{picture}(3.6,3.6)
\multiput(.3,.3)(0,1){3}{\fr}
\put(1.3,2.3){\fr}
\put(2.3,2.3){\fr}
\end{picture}}}

\newcommand{\threetwoonebox}
{\stln \lower3.8ex\hbox{
\begin{picture}(3.6,3.6)
\multiput(.3,.3)(0,1){3}{\fr}
\put(1.3,2.3){\fr}
\put(2.3,2.3){\fr}
\put(1.3,1.3){\fr}
\end{picture}}}

\newcommand{\threetwotwobox}
{\stln \lower3.8ex\hbox{
\begin{picture}(3.6,3.6)
\multiput(.3,.3)(0,1){3}{\fr}
\multiput(1.3,.3)(0,1){3}{\fr}
\put(2.3,2.3){\fr}
\end{picture}}}

\newcommand{\threethreeonebox}
{\stln \lower3.8ex\hbox{
\begin{picture}(3.6,3.6)
\multiput(.3,2.3)(1,0){3}{\fr}
\multiput(.3,1.3)(1,0){3}{\fr}
\put(.3,.3){\fr}
\end{picture}}}

\newcommand{\threethreetwobox}
{\stln \lower3.8ex\hbox{
\begin{picture}(3.6,3.6)
\multiput(.3,2.3)(1,0){3}{\fr}
\multiput(.3,1.3)(1,0){3}{\fr}
\put(.3,.3){\fr}
\put(1.3,.3){\fr}
\end{picture}}}

\newcommand{\threethreethreebox}
{\stln \lower3.8ex\hbox{
\begin{picture}(3.6,3.6)
\multiput(.3,2.3)(1,0){3}{\fr}
\multiput(.3,1.3)(1,0){3}{\fr}
\multiput(.3,.3)(1,0){3}{\fr}
\end{picture}}}

\newcommand{\gencolbox}
{\stln \lower8.6ex\hbox{
\begin{picture}(1.6,7.6)
\multiput(.3,4.3)(0,1){3}{\fr}
\put(.3,1.3){\framebox(1,3){$\vdots$}}
\put(.3,.3){\fr}
\end{picture}}}

\newcommand{\sonebox}
{\stln \lower1.4ex\hbox{
\begin{picture}(1.6,1.6)
\put(.3,.3){\sfr}
\end{picture}}}

\newcommand{\stwobox}
{\stln \lower1.4ex\hbox{
\begin{picture}(2.6,1.6)
\put(.3,.3){\sfr}
\put(1.3,.3){\sfr}
\end{picture}}}

\newcommand{\sthreebox}
{\stln \lower1.4ex\hbox{
\begin{picture}(3.6,1.6)
\multiput(.3,.3)(1,0){3}{\sfr}
\end{picture}}}

\newcommand{\sfourbox}
{\stln \lower1.4ex\hbox{
\begin{picture}(4.6,1.6)
\multiput(.3,.3)(1,0){4}{\sfr}
\end{picture}}}

\newcommand{\sfivebox}
{\stln \lower1.4ex\hbox{
\begin{picture}(5.6,1.6)
\multiput(.3,.3)(1,0){5}{\sfr}
\end{picture}}}

\newcommand{\ssixbox}
{\stln \lower1.4ex\hbox{
\begin{picture}(6.6,1.6)
\multiput(.3,.3)(1,0){6}{\sfr}
\end{picture}}}

\newcommand{\sgenrowbox}
{\stln \lower1.4ex\hbox{
\begin{picture}(7.6,1.6)
\multiput(.3,.3)(1,0){3}{\sfr}
\put(3.3,.3){\framebox(3,1){$\cdots$}}
\put(6.3,.3){\sfr}
\end{picture}}}

\newcommand{\soneonebox}
{\stln \lower2.6ex\hbox{
\begin{picture}(1.6,2.6)
\put(.3,.3){\sfr}
\put(.3,1.3){\sfr}
\end{picture}}}

\newcommand{\stwoonebox}
{\stln \lower2.6ex\hbox{
\begin{picture}(2.6,2.6)
\put(.3,1.3){\sfr}
\put(1.3,1.3){\sfr}
\put(0.3,0.3){\sfr}
\end{picture}}}

\newcommand{\sthreeonebox}
{\stln \lower2.6ex\hbox{
\begin{picture}(3.6,2.6)
\multiput(.3,1.3)(1,0){3}{\sfr}
\put(.3,.3){\sfr}
\end{picture}}}

\newcommand{\sfouronebox}
{\stln \lower2.6ex\hbox{
\begin{picture}(4.6,2.6)
\multiput(.3,1.3)(1,0){4}{\sfr}
\put(.3,.3){\sfr}
\end{picture}}}

\newcommand{\sfiveonebox}
{\stln \lower2.6ex\hbox{
\begin{picture}(5.6,2.6)
\multiput(.3,1.3)(1,0){5}{\sfr}
\put(.3,.3){\sfr}
\end{picture}}}

\newcommand{\ssixonebox}
{\stln \lower2.6ex\hbox{
\begin{picture}(6.6,2.6)
\multiput(.3,1.3)(1,0){6}{\sfr}
\put(.3,.3){\sfr}
\end{picture}}}

\newcommand{\stwotwobox}
{\stln \lower2.6ex\hbox{
\begin{picture}(2.6,2.6)
\put(.3,.3){\sfr}
\put(.3,1.3){\sfr}
\put(1.3,.3){\sfr}
\put(1.3,1.3){\sfr}
\end{picture}}}

\newcommand{\sthreetwobox}
{\stln \lower2.6ex\hbox{
\begin{picture}(3.6,2.6)
\multiput(.3,1.3)(1,0){3}{\sfr}
\put(.3,.3){\sfr}
\put(1.3,.3){\sfr}
\end{picture}}}

\newcommand{\sfourtwobox}
{\stln \lower2.6ex\hbox{
\begin{picture}(4.6,2.6)
\multiput(.3,1.3)(1,0){4}{\sfr}
\put(.3,.3){\sfr}
\put(1.3,.3){\sfr}
\end{picture}}}

\newcommand{\sfivetwobox}
{\stln \lower2.6ex\hbox{
\begin{picture}(5.6,2.6)
\multiput(.3,1.3)(1,0){5}{\sfr}
\put(.3,.3){\sfr}
\put(1.3,.3){\sfr}
\end{picture}}}

\newcommand{\ssixtwobox}
{\stln \lower2.6ex\hbox{
\begin{picture}(6.6,2.6)
\multiput(.3,1.3)(1,0){6}{\sfr}
\put(.3,.3){\sfr}
\put(1.3,.3){\sfr}
\end{picture}}}

\newcommand{\sthreethreebox}
{\stln \lower2.6ex\hbox{
\begin{picture}(3.6,2.6)
\multiput(.3,1.3)(1,0){3}{\sfr}
\multiput(.3,.3)(1,0){3}{\sfr}
\end{picture}}}

\newcommand{\sfourthreebox}
{\stln \lower2.6ex\hbox{
\begin{picture}(4.6,2.6)
\multiput(.3,1.3)(1,0){4}{\sfr}
\multiput(.3,.3)(1,0){3}{\sfr}
\end{picture}}}

\newcommand{\sfivethreebox}
{\stln \lower2.6ex\hbox{
\begin{picture}(5.6,2.6)
\multiput(.3,1.3)(1,0){5}{\sfr}
\multiput(.3,.3)(1,0){3}{\sfr}
\end{picture}}}

\newcommand{\ssixthreebox}
{\stln \lower2.6ex\hbox{
\begin{picture}(6.6,2.6)
\multiput(.3,1.3)(1,0){6}{\sfr}
\multiput(.3,.3)(1,0){3}{\sfr}
\end{picture}}}

\newcommand{\sfourfourbox}
{\stln \lower2.6ex\hbox{
\begin{picture}(4.6,2.6)
\multiput(.3,1.3)(1,0){4}{\sfr}
\multiput(.3,.3)(1,0){4}{\sfr}
\end{picture}}}

\newcommand{\sfivefourbox}
{\stln \lower2.6ex\hbox{
\begin{picture}(5.6,2.6)
\multiput(.3,1.3)(1,0){5}{\sfr}
\multiput(.3,.3)(1,0){4}{\sfr}
\end{picture}}}

\newcommand{\ssixfourbox}
{\stln \lower2.6ex\hbox{
\begin{picture}(6.6,2.6)
\multiput(.3,1.3)(1,0){6}{\sfr}
\multiput(.3,.3)(1,0){4}{\sfr}
\end{picture}}}

\newcommand{\sfivefivebox}
{\stln \lower2.6ex\hbox{
\begin{picture}(5.6,2.6)
\multiput(.3,1.3)(1,0){5}{\sfr}
\multiput(.3,.3)(1,0){5}{\sfr}
\end{picture}}}

\newcommand{\ssixfivebox}
{\stln \lower2.6ex\hbox{
\begin{picture}(6.6,2.6)
\multiput(.3,1.3)(1,0){6}{\sfr}
\multiput(.3,.3)(1,0){5}{\sfr}
\end{picture}}}

\newcommand{\ssixsixbox}
{\stln \lower2.6ex\hbox{
\begin{picture}(6.6,2.6)
\multiput(.3,1.3)(1,0){6}{\sfr}
\multiput(.3,.3)(1,0){6}{\sfr}
\end{picture}}}

\newcommand{\soneoneonebox}
{\stln \lower3.8ex\hbox{
\begin{picture}(1.6,3.6)
\multiput(.3,.3)(0,1){3}{\sfr}
\end{picture}}}

\newcommand{\stwooneonebox}
{\stln \lower3.8ex\hbox{
\begin{picture}(2.6,3.6)
\multiput(.3,.3)(0,1){3}{\sfr}
\put(1.3,2.3){\sfr}
\end{picture}}}

\newcommand{\stwotwoonebox}
{\stln \lower3.8ex\hbox{
\begin{picture}(2.6,3.6)
\multiput(.3,.3)(0,1){3}{\sfr}
\put(1.3,1.3){\sfr}
\put(1.3,2.3){\sfr}
\end{picture}}}

\newcommand{\stwotwotwobox}
{\stln \lower3.8ex\hbox{
\begin{picture}(2.6,3.6)
\multiput(.3,.3)(0,1){3}{\sfr}
\multiput(1.3,.3)(0,1){3}{\sfr}
\end{picture}}}

\newcommand{\sthreeoneonebox}
{\stln \lower3.8ex\hbox{
\begin{picture}(3.6,3.6)
\multiput(.3,.3)(0,1){3}{\sfr}
\put(1.3,2.3){\sfr}
\put(2.3,2.3){\sfr}
\end{picture}}}

\newcommand{\sthreetwoonebox}
{\stln \lower3.8ex\hbox{
\begin{picture}(3.6,3.6)
\multiput(.3,.3)(0,1){3}{\sfr}
\put(1.3,2.3){\sfr}
\put(2.3,2.3){\sfr}
\put(1.3,1.3){\sfr}
\end{picture}}}

\newcommand{\sthreetwotwobox}
{\stln \lower3.8ex\hbox{
\begin{picture}(3.6,3.6)
\multiput(.3,.3)(0,1){3}{\sfr}
\multiput(1.3,.3)(0,1){3}{\sfr}
\put(2.3,2.3){\sfr}
\end{picture}}}

\newcommand{\sthreethreeonebox}
{\stln \lower3.8ex\hbox{
\begin{picture}(3.6,3.6)
\multiput(.3,2.3)(1,0){3}{\sfr}
\multiput(.3,1.3)(1,0){3}{\sfr}
\put(.3,.3){\sfr}
\end{picture}}}

\newcommand{\sthreethreetwobox}
{\stln \lower3.8ex\hbox{
\begin{picture}(3.6,3.6)
\multiput(.3,2.3)(1,0){3}{\sfr}
\multiput(.3,1.3)(1,0){3}{\sfr}
\put(.3,.3){\sfr}
\put(1.3,.3){\sfr}
\end{picture}}}

\newcommand{\sthreethreethreebox}
{\stln \lower3.8ex\hbox{
\begin{picture}(3.6,3.6)
\multiput(.3,2.3)(1,0){3}{\sfr}
\multiput(.3,1.3)(1,0){3}{\sfr}
\multiput(.3,.3)(1,0){3}{\sfr}
\end{picture}}}

\newcommand{\sgencolbox}
{\stln \lower8.6ex\hbox{
\begin{picture}(1.6,7.6)
\multiput(.3,4.3)(0,1){3}{\sfr}
\put(.3,1.3){\framebox(1,3){$\vdots$}}
\put(.3,.3){\sfr}
\end{picture}}}

\newcommand{\sgenrowonebox}
{\stln \lower2.6ex\hbox{
\begin{picture}(7.6,2.6)
\multiput(.3,1.3)(1,0){3}{\sfr}
\put(3.3,1.3){\framebox(3,1){$\cdots$}}
\put(6.3,1.3){\sfr}
\put(.3,.3){\sfr}
\end{picture}}}

\newcommand{\genrowonebox}
{\stln \lower2.6ex\hbox{
\begin{picture}(7.6,2.6)
\multiput(.3,1.3)(1,0){3}{\fr}
\put(3.3,1.3){\framebox(3,1){$\cdots$}}
\put(6.3,1.3){\fr}
\put(.3,.3){\fr}
\end{picture}}}

\newcommand{\sgenrowtwobox}
{\stln \lower2.6ex\hbox{
\begin{picture}(7.6,2.6)
\multiput(.3,1.3)(1,0){3}{\sfr}
\put(3.3,1.3){\framebox(3,1){$\cdots$}}

\put(6.3,1.3){\sfr}
\put(.3,.3){\sfr}
\put(1.3,.3){\sfr}
\end{picture}}}

\newcommand{\sgentworowonerowbox}
{\stln \lower2.6ex\hbox{
\begin{picture}(11.6,2.6)
\multiput(.3,1.3)(1,0){3}{\sfr}
\put(3.3,1.3){\framebox(3,1){$\cdots$}}
\put(6.3,1.3){\sfr}
\put(7.3,1.3){\framebox(3,1){$\cdots$}}
\put(10.3,1.3){\sfr}
\multiput(.3,.3)(1,0){3}{\sfr}
\put(3.3,.3){\framebox(3,1){$\cdots$}}
\put(6.3,.3){\sfr}
\end{picture}}}

\newcommand{\soneoneoneonebox}
{\stln \lower5ex\hbox{
\begin{picture}(1.6,4.6)
\multiput(.3,.3)(0,1){4}{\sfr}
\end{picture}}}

\newcommand{\soneoneoneoneoneonebox}
{\stln \lower7.4ex\hbox{
\begin{picture}(1.6,6.6)
\multiput(.3,.3)(0,1){6}{\sfr}
\end{picture}}}

\newcommand{\stwotwooneonebox}
{\stln \lower5ex\hbox{
\begin{picture}(2.6,4.6)
\multiput(.3,.3)(0,1){4}{\sfr}
\put(1.3,2.3){\sfr}
\put(1.3,3.3){\sfr}
\end{picture}}}

\newcommand{\sgenrowrowbox}
{\stln \lower2.6ex\hbox{
\begin{picture}(7.6,2.6)
\multiput(.3,1.3)(1,0){3}{\sfr}
\put(3.3,1.3){\framebox(3,1){$\cdots$}}
\put(6.3,1.3){\sfr}
\multiput(.3,.3)(1,0){3}{\sfr}
\put(3.3,.3){\framebox(3,1){$\cdots$}}
\put(6.3,.3){\sfr}
\end{picture}}}
\numberwithin{equation}{section}
\begin{document}

\begin{titlepage}
\begin{center}
\begin{large}
\textbf{Minimal unitary representation of $D(2,1;\lambda)$ and its $SU(2)$ deformations and $d=1$, $N=4$ superconformal models}
\end{large}
\vspace{1cm}
\\
\begin{large}
Karan Govil\footnote{kgovil@phys.psu.edu} and
Murat Gunaydin\footnote{murat@phys.psu.edu}
\end{large}
\\
%\end{center}
\vspace{.35cm}
 \emph{Institute for Gravitation and the Cosmos \\
 Physics Department \\
Pennsylvania State University\\
University Park, PA 16802, USA} \\
%\today
%\maketitle
\end{center}
\vspace{1cm}
\begin{abstract}
\noindent
Quantization of the geometric quasiconformal realizations of noncompact groups and supergroups leads directly to their minimal unitary representations (minreps). Using quasiconformal methods massless unitary supermultiplets of  superconformal groups $SU(2,2|N)$ and $OSp(8^*|2n)$  in four and six dimensions were constructed as minreps  and their $U(1) $ and $SU(2)$ deformations, respectively.  In this paper we extend  these results to $SU(2)$ deformations of the minrep of $N=4$ superconformal algebra $D(2,1;\lambda)$ in one dimension.   We find that $SU(2)$ deformations can be achieved using $n$ pairs of bosons and $m$ pairs of fermions simultaneously. The generators of deformed minimal representations  of $D(2,1;\lambda)$ commute with the generators of a dual superalgebra $OSp(2n^*|2m)$  realized in terms of these bosons and fermions. We show that  there exists a precise mapping between symmetry generators of $N=4$ superconformal models in harmonic superspace studied recently and minimal  unitary supermultiplets of $D(2,1;\lambda)$ deformed by a pair of bosons. This can be understood as a particular case of a general mapping between  the spectra of quantum mechanical quaternionic K\"{a}hler sigma models with eight super symmetries and minreps of their isometry groups  that descends from the precise mapping established between the  $4d$, $N=2$ sigma models coupled to supergravity and minreps of their isometry groups.
\end{abstract}
\end{titlepage}

\section{Introduction}
	
The concept of a minimal unitary representation of a noncompact Lie group was introduced by Joseph  in \cite{MR0342049}. His work was inspired by the work of physicists on spectrum generating symmetries in the 1960s.
They are defined as   unitary
representations over  Hilbert spaces
of functions of  smallest possible (minimal) number of variables.    Joseph determined  their dimensions and gave
minimal realizations of the complex forms of classical Lie algebras
and of the exceptional Lie algebra $\mathfrak{g}_2$ in a Cartan-Weyl basis.   Over the intervening decades much research  was done by the mathematicians on minimal unitary representations of Lie groups \cite{MR644845,MR1103588,MR1159103,MR1372999,MR1278630,MR1327538,MR1108044,MR2020550,MR2020551,MR2020552,Gover:2009vc}.
The minimal unitary representations of simply laced
groups studied in \cite{MR1159103} was reformulated with a view towards their applications to physics by the authors of \cite{Kazhdan:2001nx} who  also gave
their spherical vectors. 

Over the past decade,  there has been a  considerable amount of research done on the  unitary representations  of global U-duality groups of extended supergravity theories by physicists.This was partly motivated by the  proposals that certain extensions of U-duality groups must act as their spectrum generating symmetry groups.
The study of the 
orbits of extremal black hole solutions in $5d$, $N=8$ supergravity and
$N=2$ Maxwell-Einstein supergravity theories with symmetric scalar
manifolds led to the proposal that   U-duality groups of corresponding four dimensional theories must act
as spectrum generating conformal symmetry groups
\cite{Ferrara:1997uz,Gunaydin:2000xr,Gunaydin:2004ku,Gunaydin:2003qm,Gunaydin:2005gd,Gunaydin:2009pk}.  Extension of the above proposal to the   extremal black hole solutions of four dimensional supergravity theories with symmetric scalar manifolds led  to the discovery of novel geometric quasiconformal realizations  of three dimensional U-duality groups \cite{Gunaydin:2000xr}. The quasiconformal extensions of  U-duality groups were then proposed   as spectrum generating symmetry groups of the
corresponding $4d$ supergravity theories 
\cite{Gunaydin:2000xr,Gunaydin:2004ku,Gunaydin:2003qm,Gunaydin:2005gd,Gunaydin:2009pk}.  The proposal that three
dimensional U-duality groups must act as spectrum generating quasiconformal
groups was given a quantum  realization in
\cite{Gunaydin:2005mx,Gunaydin:2007bg,Gunaydin:2007qq} using the equivalence of attractor equations  of spherically
symmetric stationary BPS black holes of $4d$  supergravity
theories and the geodesic equations of a fiducial particle moving in the target space of $3d$  supergravity theories obtained by their reduction on a timelike circle \cite{Breitenlohner:1987dg}.

Quasiconformal
realization of three dimensional  U-duality group $E_{8(8)}$ of maximal supergravity
in three dimensions is the first known geometric realization of
$E_{8(8)}$\cite{Gunaydin:2000xr} and   leaves invariant
a generalized light-cone with respect to a quartic distance function  in 57
dimensions. Quasiconformal realizations exist for different  real forms of all noncompact groups as well as for their complex forms  \cite{Gunaydin:2000xr,Gunaydin:2005zz}.

The quantization of  geometric quasiconformal action of a noncompact group leads directly to its minimal unitary representation as was first shown  explicitly for the  
exceptional group  $E_{8(8)}$ with the maximal compact subgroup $SO(16)$
\cite{Gunaydin:2001bt}.
Quasiconformal construction of minimal unitary representations of  U-duality groups $\mathrm{F}_{4(4)}$, $\mathrm{E}_{6(2)}$,
 $\mathrm{E}_{7(-5)}$ ,  $\mathrm{E}_{8(-24)}$ and $SO(d+2,4)$ of  $3d$ $N=2$ Maxwell-Einstein supergravity theories with symmetric scalar manifolds were given  in
\cite{Gunaydin:2005zz,Gunaydin:2004md}.
In \cite{Gunaydin:2006vz},   a unified approach to quasiconformal 
construction  of the minimal unitary representations of 
noncompact groups was formulated 
and  extended to define and construct 
minimal unitary representations of non-compact supergroups $G$ whose even
subgroups are of the form $H\times SL(2,\mathbb{R})$ with $H$
compact.

The term minimal unitary representation refers, in general, to a unique  representation of the respective noncompact group. The symplectic group $Sp(2N,\mathbb{R})$ admits two singleton irreps with the same eigenvalue of  quadratic Casimir operator. Both of these singletons are minimal unitary representations, notwithstanding the fact that in some of the mathematics literature only the scalar singleton is referred to as the minrep.   Similarly the supergroups $OSp(M|2N,\mathbb{R}) $ with the even subgroup $SO(M) \times Sp(2N,\mathbb{R})$ admit two inequivalent singleton supermultiplets \cite{Gunaydin:1985tc,Gunaydin:1988kz,Gunaydin:1987hb} which should both be interpreted as minimal unitary supermultiplets.

However, for noncompact groups or supergroups that do not admit singleton irreps but infinitely many  doubleton irreps,  this raises the question as to whether any of the doubleton irreps can be identified with the  minimal representation. If so  how  is then the remaing infinite set of doubletons  related to the minrep?
%%%%%%%%%%%%%%TBC%%%%%%%%%%%%%%
Authors of  \cite{Fernando:2009fq} investigated this question for  $4D$ conformal group $SU(2,2)$ and   supergroups $SU(2,2|N)$  and showed that
 the minimal unitary representation of $SU(2,2)$   is simply  the scalar doubleton representation corresponding to a massless scalar field in four dimensions. Furthermore they  showed that the minrep of $SU(2,2)$ admits a one-parameter family ($\zeta$) of deformations. For a positive
(negative)  integer value of the deformation parameter $\zeta$, quasiconformal approach leads to
a positive energy
unitary irreducible representation  corresponding to a
massless conformal field in four dimensions  transforming in  $\left( 0
\,,\, \frac{\zeta}{2} \right)$ $\left( \left( -\frac{\zeta}{2} \,,\, 0
\right) \right)$ representation of the Lorentz subgroup, $SL(2,\mathbb{C})$ of $SU(2,2)$. These ``deformed minimal unitary representations" are simply the doubleton representations of $SU(2,2)$  corresponding to massless conformal fields in four dimensions \cite{Gunaydin:1998sw,Gunaydin:1998jc}. 
These results extend to the minimal unitary representations of supergroups $SU(2,2\,|\,N)$ with the even subgroups $SU(2,2)\times U(N)$  and their deformations as well as to more general supergroups $SU(m,n|N)$. 
The minimal unitary supermultiplet of $SU(2,2|N)$ is  the CPT self-conjugate  doubleton supermultiplet, and for $PSU(2,2|4)$ it is simply the four dimensional $N=4$  Yang-Mills supermultiplet  \cite{Fernando:2009fq}.

Similar results were obtained for the $6d$ conformal group $SO^*(8)$ and its supersymmetric extensions $OSp(8^*|2N)$ in \cite{Fernando:2010dp,Fernando:2010ia}. In the case of 
$SO^*(8)$ and $OSp(8^*|2N)$ the deformations of the minrep are labeled by the eigenvalues of Casimir of an $SU(2)$ sub algebra. Minimal unitary supermultiplet of $OSp(8^*|4)$ turns out to be the $(2,0)$ conformal supermultiplet whose field theory was predicted to live on the boundary of $AdS_7$ as conformally invariant theory\cite{Gunaydin:1984wc} and whose interacting theory is believed to be dual to M-theory on $AdS_7\times S^4$ \cite{Maldacena:1997re}. The deformed minimal supermultiplets of $OSp(8^*|4)$ are simply the doubleton 
supermultiplets studied in  \cite{Gunaydin:1999ci,Fernando:2001ak} .

As was shown in  \cite{Gunaydin:2007vc}  that there exists 
 a remarkable connection between the harmonic superspace (HSS) formulation of $4d$ , $N=2$ supersymmetric quaternionic K\"{a}hler sigma models that couple to $N=2$ supergravity and the minimal unitary representations of their isometry groups obtained by quasiconformal methods. In particular, for $N=2$ sigma models with quaternionic symmetric target spaces of the form $ \frac{G}{H\times SU(2)}$  one finds   a one-to-one mapping between the Killing potentials that generate the isometry group $G$ under poisson brackets  in the HSS formulation and the generators of  minimal unitary representation of G obtained from its quasiconformal realization. 
 What this implies is that the fundamental quantum spectra of these sigma models must furnish minimal unitary representation of the isometry group and the full spectrum is obtained by tensoring of the minrep.\footnote{ For a free theory the fundamental spectrum is simply the Fock space of the oscillators corresponding to the Fourier modes of the free fields.} 
 Since the quantization of $4D$ sigma models is problematic and their quantum completion may require extension to superstring theory, it was suggested that they be dimensionally reduced to lower dimensions and quantized so as to test the above proposal. In particular it was predicted that fundamental spectra of quantum mechanical models with 8 super symmetries obtained  by reduction to one dimension must furnish the minimal unitary representations of their global symmetry groups \cite{Gunaydin:2007vc,Gunaydin:2009pk}. 

In this paper we study the deformations of the minimal unitary supermultiplet of $D(2,1;\lambda)$ with the even subgroup $SU(1,1)\times SU(2)\times SU(2)$ . The motivations for  this study  are manifold. $D(2,1;\lambda)$  represents a one parameter family of $N=4$ conformal algebras in one dimension and  is relevant to $AdS_2/CFT_1$ dualities. It is  also important for $AdS_3/CFT_2$ dualities since the $AdS_3$ group $SO(2,2)$ factorizes as $SU(1,1)\times SU(1,1)$. Supersymmetric extensions factorize as well and each factor can be extended to $D(2,1;\lambda)$ \cite{Gunaydin:1986fe}. 
Another motivation is to investigate the connection between the spectra of $N=4$ superconformal quantum mechanical models that have been studied in recent years \cite{Galajinsky:2003cw,Ivanov:2003nk,Ivanov:2002pc,Krivonos:2010zy,Lechtenfeld:2010sk,Fedoruk:2010wj,Fedoruk:2009xf,Fedoruk:2009pq,Delduc:2011gt,
Ivanov:2011gk,Ivanov:2009hca,Fedoruk:2008hk,Delduc:2008ca,Delduc:2006yp,Delduc:2006pg,Galajinsky:2007gm,Galajinsky:2008nj,
Bellucci:2010te,Bellucci:2010uq,Bellucci:2009dz,Bellucci:2008fx,Bellucci:2006ts,Fedoruk:2011aa} and the deformations of minimal unitary supermultiplets of corresponding conformal superalgebras in the light of the results of \cite{Gunaydin:2007vc,Gunaydin:2009pk}. 

The plan of the paper is as follows. In section 2 we review the minimal unitary realization of $D(2,1;\lambda)$ as constructed in \cite{Gunaydin:2006vz} using quasiconformal techniques. In section 3 we construct the $SU(2)$ deformations of the minrep of $D(2,1;\lambda)$ using bosonic oscillators in the  noncompact 5-graded basis. In Section 4 we reformulate the results of section 3 in the compact 3-graded basis and  show that the deformations of the minrep of $D(2,1;\lambda)$ are  positive ``energy" ( unitary lowest weight ) representations of  $D(2,1;\lambda)$. We then present the corresponding unitary supermultiplets.  Section 5 discusses 
deformations of the minrep using both bosons and fermions and how the deformed $D(2,1;\lambda)$ commutes with a noncompact super algebra $OSp(2n^*|2m)$ with the even subgroup $SO^*(2m)\times USp(2n) $ constructed using ``deformation" bosons and fermions. In section 6 we review some of the  results of work on $N=4$ superconformal mechanics  and show how its  symmetry generators and spectrum map into the generators of $D(2,1;\lambda)$  deformed by a pair of bosonic oscillators and the resulting unitary supermultiplets. We conclude with a brief discussion of our results.

\section{The minimal unitary representation  of $D(2,1;\lambda)$}

Of all the noncompact real forms of the one parameter family of  supergroups  $D(2,1;\lambda)$ only the real form with the even subgroup $SU(2)\times SU(2) \times SU(1,1)$ admits unitary lowest weight ( positive energy ) representations. In this paper we shall study the minimal unitary representations of this real form, which we shall  denote as $D(2,1;\lambda)$ ,  and their deformations. 
We shall label the even subgroup as $SU(2)_A\times SU(2)_T \times SU(1,1)_K$ with the odd generators transforming in the $(1/2,1/2,1/2)$ representation with respect to it.

The Lie super algebra of $D(2,1;\lambda)$ can be given a 5-graded decomposition of the form
\be
D(2,1;\lambda) = \mathfrak{g}^{(-2)} \oplus \mathfrak{g}^{(-1)} \oplus \mathfrak{g}^{(0)}	\oplus \mathfrak{g}^{(+1)}	\oplus \mathfrak{g}^{(+2)}
\ee
where the grade $\pm 2$ subspaces are one dimensional and the grade zero sub algebra is 
\be
\mathfrak{g}^{(0)} = \mathfrak{su}(2)_A \oplus \mathfrak{su}(2)_T \oplus \mathfrak{so}(1,1)_{\Delta}
\ee

The grade $\pm 1$ subspaces contain 4 odd generators transforming in the $(1/2,1/2)$ representation of $SU(2)_A\times SU(2)_T$ subgroup and the generators  belonging to  $\mathfrak{g}^{(-2)}$ and  $\mathfrak{g}^{(+2)}$ together with the $SO(1,1)$  generator $\Delta$ of grade zero subspace form the $\mathfrak{su}(1,1)_K$ subalgebra.

For $\lambda=-1/2$ the Lie superalgebra $D(2,1;\lambda)$ is isomorphic to $OSp(4^*|2)=OSp(4|2,\mathbb{R})$ whose representations were studied in \cite{Gunaydin:1990ag, Gunaydin:1988kz} using the twistorial oscillator methods. The minrep of $D(2,1;\lambda)$ was obtained  in \cite{Gunaydin:2006vz} using quasiconformal techniques. We shall reformulate the minrep given in \cite{Gunaydin:2006vz} and  study its deformations as was done for 4d and 6d superconformal algebras \cite{Fernando:2009fq,Fernando:2010ia,Fernando:2010dp}. 
%%%%%%%%%%%%%%%%%%%%%%%%%%%%%%
The minimal unitary representation of $D(2,1;\lambda)$ in the Hilbert space of a single bosonic coordinate and four fermionic coordinates was given in \cite{Gunaydin:2006vz}. The generators belonging to various grade   
subspaces were labelled  with respect to the 
$SU(2)_A\times SU(2)_T \times SO(1,1)_\Delta$ subgroup  as follows
\begin{equation}
D\left(2,1;\lambda \right) = \mathbf{(0,0)}^{-2} \oplus \left(
\mathbf{1/2},\mathbf{1/2} \right)^{-1} \oplus
  \left( \mathfrak{su}\left(2\right)_A \oplus \mathfrak{su}\left(2\right)_T \oplus \Delta \right)
  \oplus \left( \mathbf{1/2},\mathbf{1/2} \right)^{+1} \oplus  \mathbf{(0,0)}^{+2}
\end{equation}
\be
D\left(2,1;\sigma\right) = E \oplus E^{\alpha,\Dot{\alpha}} \oplus \left(M_{(1)}^{\alpha,\beta} + M_{(2)}^{\Dot{\alpha},\Dot{\beta}} +\Delta \right) \oplus F^{\alpha,\dot{\alpha}} \oplus F 
\ee
The  single bosonic coordinate and its canonical momentum $(x, p)$ satisfy
\be
[ x, p] =i
\ee
The four fermionic ``coordinates"  $X^{\alpha, \Dot{\alpha}}$  satisfy the  anti-commutation relations \cite{Gunaydin:2006vz}:
\begin{equation}
   \left\{ X^{\alpha,\Dot{\alpha}}, X^{\beta,\Dot{\beta}} \right\} = \epsilon^{\alpha\beta} \epsilon^{\Dot{\alpha}\Dot{\beta}}
\end{equation}
where $\alpha, \Dot{\alpha},..$ denote the spinor indices of 
$SU(2)_A $ and $ SU(2)_T$ subgroups and $\epsilon_{\alpha\beta}$ and $\epsilon_{\dot{\alpha}\dot{\beta}}$ are the Levi-Civita tensors in the respective spaces .  The generators belonging to the negative and zero grade subspaces are realized as bilinears  
\begin{equation}
  E = \frac{1}{2} x^2 \qquad E^{\alpha,\Dot{\alpha}} = x X^{\alpha,\Dot{\alpha}} \qquad
   \Delta = \frac{1}{2} \left(x p + p x \right)
\end{equation}
\begin{equation}
\begin{split}
   M_{(1)}^{\alpha,\beta} &= \frac{1}{4} \epsilon_{\Dot{\alpha}\Dot{\beta}}
    \left( X^{\alpha,\Dot{\alpha}} X^{\beta,\Dot{\beta}} - X^{\beta,\Dot{\beta}} X^{\alpha,\Dot{\alpha}}  \right) \\
   M_{(2)}^{\Dot{\alpha},\Dot{\beta}} &= \frac{1}{4} \epsilon_{\alpha\beta}
    \left( X^{\alpha,\Dot{\alpha}} X^{\beta,\Dot{\beta}} - X^{\beta,\Dot{\beta}} X^{\alpha,\Dot{\alpha}}  \right)
\end{split}
\end{equation}
They satisfy the (super)commutation relations 
\begin{equation}
\begin{split}
   \left[ M^{\alpha,\beta}_{(1)}, M^{\lambda,\mu}_{(1)} \right] &=
       \epsilon^{\lambda\beta} M^{\alpha,\mu}_{(1)} + \epsilon^{\mu\alpha} M^{\beta,\lambda}_{(1)} \\
   \left[ M^{\Dot{\alpha},\Dot{\beta}}_{(2)}, M^{\Dot{\lambda},\Dot{\mu}}_{(2)} \right] &=
       \epsilon^{\Dot{\lambda}\Dot{\beta}} M^{\Dot{\alpha},\Dot{\mu}}_{(2)} +
       \epsilon^{\Dot{\mu}\Dot{\alpha}} M^{\Dot{\beta},\Dot{\lambda}}_{(2)} \\
   \left[ M^{\alpha,\beta}_{(1)}, M^{\Dot{\lambda},\Dot{\mu}}_{(2)} \right] &= 0 \\
\left\{E^{\alpha,\Dot{\alpha}}, E^{\beta,\Dot{\beta}} \right\} & = 2 \, \epsilon^{\alpha\beta} \epsilon^{\Dot{\alpha}\Dot{\beta}}
\, E
\end{split}
\end{equation}
The quadratic Casimirs of the two $SU(2)$s are 
\begin{equation}
   \mathcal{I}_4 = \epsilon_{\alpha\beta}\epsilon_{\lambda\mu} M_{(1)}^{\alpha\lambda} M_{(1)}^{\beta\mu}
  \qquad
   \mathcal{J}_4 = \epsilon_{\Dot{\alpha}\Dot{\beta}}\epsilon_{\Dot{\lambda}\Dot{\mu}}
         M_{(2)}^{\Dot{\alpha}\Dot{\lambda}} M_{(2)}^{\Dot{\beta}\Dot{\mu}}
\end{equation}
and differ by  a c-number $\mathcal{I}_4 + \mathcal{J}_4 =
-\frac{3}{2}$. Hence we can  use either one of them to express the  generator $F$  of
$\mathfrak{g}^{+2}$ subspace as 
\begin{equation}
   F = \frac{1}{2} p^2 + \frac{\sigma}{x^2} \left( \mathcal{I}_4 + \frac{3}{4} + \frac{9}{8} \sigma \right)
\end{equation}
The grade +1 generators are then given by 
\begin{equation}
   F^{\alpha\Dot{\alpha}} = -i \left[ E^{\alpha\Dot{\alpha}}, F \right]
\end{equation}
and  one finds 
\begin{equation}
  \left\{ F^{\alpha\Dot{\alpha}}, F^{\beta\Dot{\beta}} \right\} = 2 \epsilon^{\alpha\beta} \epsilon^{\Dot{\alpha}\Dot{\beta}} F
  \qquad \left[ F^{\alpha\Dot{\alpha}}, F \right ] = 0
\end{equation}
and 
\begin{equation}
  \left\{ F^{\alpha\Dot{\alpha}}, E^{\beta\Dot{\beta}} \right\} =
       \epsilon^{\alpha\beta} \epsilon^{\Dot{\alpha}\Dot{\beta}} \Delta -
       \left( 1- 3 \sigma \right) i \epsilon^{\alpha\beta} M_{(2)}^{\Dot{\alpha}\Dot{\beta}} -
       \left( 1+ 3 \sigma \right) i \epsilon^{\Dot{\alpha}\Dot{\beta}} M_{(1)}^{\alpha\beta}
\end{equation}
 For $\sigma=0$ the superalgebra
$D\left(2,1,\sigma\right)$ is isomorphic to $OSp(4 \vert
2,\mathbb{R})$ and for the values $\sigma= \pm \frac{1}{3}$ it
reduces to
\begin{equation*}
SU\left(2 \vert 1,1\right)\times SU\left(2\right)
\end{equation*}

In this paper we shall use a different label $\lambda$ for  the one parameter superalgebras
$D(2,1;\lambda)$ which is  related to the label $\sigma$ as:
\be 
\sigma = \frac{2\lambda +1}{3}
\ee 
With this labeling we have
\be
D(2,1; \lambda= -1/2) = OSp(4|2,\mathbb{R})
\ee

%%%%%%%%%%%%%%%%%%%%%%%%%%%%%%
\section{$SU(2)$ Deformations of the minimal unitary supermultiplet of $D(2,1;\lambda)$ }
\label{sect-construction}
To deform the minrep  of $D(2,1;\lambda)$ given above we shall first rewrite the (super)commutation relations of its generators in a split basis in which the $U(1)$ generators of the two $SU(2)$ subgroups  are diagonalized 
\bea
SU(2)_A  &\Longrightarrow \,\, A_+ , A_-, A_0 \\
SU(2)_T & \Longrightarrow \,\, T_+, T_- , T_0 \\
%SU(1,1)_K &\Longrightarrow  \,\, K_+, K_- , K_0 
\eea
and the fermionic ``coordinates" $X^{\alpha,\Dot{\alpha}}$ are written as fermionic annihilation and creation operators $\alpha (\alpha^\dagger)$ and $\beta (\beta^\dagger)$ with definite values of $U(1)$ charges
\be 
\{ \alpha , \alpha^\dagger \} =1 = \{ \beta , \beta^\dagger \} 
\ee
\be \{ \alpha , \beta^\dagger \} =0= \{ \beta , \alpha^\dagger \}
\ee
We shall choose the fermionic Fock vacuum such that
\be
\alpha |0\rangle_F=0 =\beta|0\rangle_F
\ee
The generators of $SU(2)_T$ are then given by the following bilinears of these
fermionic oscillators:
\begin{equation}
T_+ = \alpha^\dagger \beta
\qquad \qquad
T_- = \beta^\dagger \alpha
\qquad \qquad
T_0 = \frac{1}{2} \left( N_\alpha - N_\beta  \right)
\label{SU(2)G_gen}
\end{equation}
where $N_\alpha = \alpha^\dagger \alpha$ and $N_\beta = \beta^\dagger \beta$ are the respective
number operators. They satisfy the commutation relations:
\begin{equation}
\commute{T_+}{T_-} = 2 \, T_0
\qquad \qquad \qquad
\commute{T_0}{T_\pm} = \pm T_\pm
\end{equation}
with the Casimir
\be T^2
= {T_0}^2 + \frac{1}{2} \left( T_+ T_- + T_- T_+ \right) \ee 
The generators of the subalgebra $\mathfrak{su}(2)_A$  are given by the following bilinears of fermionic oscillators:
\begin{equation}
\begin{aligned}
A_+ &= \alpha^\dagger \beta^\dagger
\\
A_- &= \left( A_+ \right)^\dag = \beta \alpha 
\\
A_0 &= \frac{1}{2} \left( N_\alpha + N_\beta-1 \right)
\end{aligned}
\end{equation}
The quadratic Casimir of the subalgebra $\mathfrak{su}(2)_A$ is
\begin{equation}
\mathcal{C}_2 \left[ \mathfrak{su}(2)_A \right]
= A^2
= {A_0}^2 + \frac{1}{2} \left( A_+ A_- + A_- A_+ \right) \,.
\end{equation}
and is related to $T^2$ as follows 
\be 
T^2  + A^2= \frac{3}{4} 
\ee 
The four states in the Fock space of two fermions ($\alpha, \, \beta$) 
transform in the $(1/2,1/2)$ representation of $SU(2)_A \times SU(2)_T$, We shall label them  by their eigenvalues under $T_0$ and $A_0$ as $|m_t;m_a \rangle$:
\bea
T_0 |m_t;m_a \rangle = m_t |m_t;m_a \rangle \\
A_0|m_t;m_a \rangle = m_a |m_t;m_a \rangle 
\eea
More explicitly we have
\[
\begin{array}{ccccc}
\ket{0}_F \eq \ket{0,-\half}_F \eq \ket{0,\dowa}_F \\
\alpha^\dagger \beta^\dagger \ket{0}_F \eq \ket{0,\half,}_F \eq \ket{0,\upa}_F \\
\alpha^\dagger\ket{0}_F \eq \ket{\half,0}_F \eq |\upa,0\rangle_F \\
\beta^\dagger\ket{0}_F \eq \ket{-\half,0}_F \eq |\dowa,0\rangle_F 
\end{array}
\]
where $\upa$ denotes $+1/2$ and $\dowa$ denotes $-1/2$ eigenvalue of the respective $U(1)$ generator.

The grade -1 generators can then be written as bilinears of the coordinate $x$ with the fermionic oscillators:

\begin{equation}
\begin{aligned}
Q = x \, \alpha
\\
S = x \, \beta
\end{aligned}
\qquad \qquad \qquad \qquad
\begin{aligned}
Q^\dagger = x \, \alpha^\dagger
\\
S^\dagger = x \, \beta^\dagger
\end{aligned}
\end{equation}
They close into $K_-$ under anti-commutation:
\[
 \{ Q , Q^\dagger \} = 2K_-
\]
The grade zero generators in the five grading determined by $\Delta$ are the generators $T_+ , T_-$ and $T_0$ of $SU(2)_T$ , $A_+ , A_-$ and $A_0$ of $SU(2)_A$ and $\Delta$.
The  grade $+2$ generator with respect to $\Delta$ is given by:
\bea
K_+ &=& \frac{1}{2}p^2+\frac{1}{x^2}\left(2\lambda T^2+\frac{2}{3}(\lambda-1)A^2+\frac{3}{8}+\frac{1}{2} \lambda(\lambda-1)\right) \nonumber \\
K_+ &=& \frac{1}{2}p^2+\frac{1}{x^2}\left(\frac{2}{3} ( 2 \lambda +1)  T^2+ \frac{\lambda^2}{2} +1 \right) 
\eea

%%%%%%%%%%%%%%%%%%%%TBC%%%%%%%%%%%%%%%%
%%%%%%%%%%%%%%%%%%%%%%%%%%%%%%%%%%%%%%%

 In order to obtain unitary irreducible  representations that are ``deformations" of the minrep of $D(2,1;\lambda)$ we introduce bosonic oscillators
$a_m$, $b_m$ and their hermitian conjugates $a^m = \left( a_m \right)^\dag$,
$b^m = \left( b_m \right)^\dag$ ($m,n,\dots = 1,2$) that satisfy the
commutation relations:
\begin{equation}
\commute{a_m}{a^n}
= \commute{b_m}{b^n}
= \delta^n_m
\qquad \qquad
\commute{a_m}{a_n}
= \commute{a_m}{b_n}
= \commute{b_m}{b_n}
= 0
\end{equation}
and introduce an $SU(2)_S$ Lie algebra whose generators are as follows:
\begin{equation}
S_+ = a^m b_m
\qquad \qquad
S_- = \left( S_+ \right)^\dag
    = a_m b^m
\qquad \qquad
S_0 = \frac{1}{2} \left( N_a - N_b \right)
\label{SU(2)S_gen}
\end{equation}
where $N_a = a^m a_m$ and $N_b = b^m b_m$ are the respective number
operators. They satisfy:
\begin{equation}
\commute{S_+}{S_-} = 2 \, S_0
\qquad \qquad \qquad
\commute{S_0}{S_\pm} = \pm S_\pm
\end{equation}
The quadratic Casimir of $\mathfrak{su}(2)_S$ is
\begin{equation}
\begin{split}
\mathcal{C}_2 \left[ \mathfrak{su}(2)_S \right]
 = S^2
&= {S_0}^2 + \frac{1}{2} \left( S_+ S_- + S_- S_+ \right)
\\
&= \frac{1}{2} \left(N_a + N_b \right)
   \left[ \frac{1}{2} \left( N_a + N_b \right) + 1 \right]
   - 2 a^{[m} b^{n]} \, a_{[m} b_{n]}
\end{split}
\end{equation}
where square bracketing $a_{[m} b_{n]} = \frac{1}{2} \left( a_m b_n - a_n b_m
\right)$ represents antisymmetrization of weight one.
The bilinears $a_{[m} b_{n]}$ and $a^{[m} b^{n]}$  close into 
 $U(P)$ generated by the bilinears
\be
U^m_{~ n} = a^m a_n + b^m b_n 
\ee
under commutation and all together they form the Lie algebra of noncompact group $SO^*(2P)$ with the maximal compact subgroup $U(P)$. The group $SO^*(2P)$ thus generated 
 commutes with  $SU(2)_S$  as well as with $D(2,1;\lambda)$. 
 
To obtain the $SU(2)$ deformed $D(2,1;\lambda)$ superalgebra
we replace  the generators of $\mathfrak{su}(2)_T$  subalgebra with the   generators of the diagonal subgroup of $SU(2)_T $ and $SU(2)_S$  
\be
\mathfrak{su}(2)_T  \Longrightarrow  \mathfrak{su}(2)_S \oplus \mathfrak{su}(2)_T
\ee
and denote the diagonal subgroup as $SU(2)_{\mathcal{T}}$  and its Lie algebra as $\mathfrak{su}(2)_{\mathcal{T}} $. Its generators  are simply:
\begin{equation}
\begin{split}
\mathcal{T}_+
&= S_+ + T_+
 = a^m b_m + \alpha^\dagger \beta
\\
\mathcal{T}_-
&= S_- + T_-
 = b^m a_m + \beta^\dagger \alpha
\\
\mathcal{T}_0
&= S_0 + T_0
 = \frac{1}{2} \left( N_a - N_b + N_\alpha - N_\beta \right)
\end{split}
\label{SU(2)T_gen}
\end{equation}
with the quadratic Casimir 
\begin{equation}
\mathcal{C}_2 \left[ \mathfrak{su}(2)_T \right]
= T^2
= {\mathcal{T}_0}^2 + \frac{1}{2} \left( \mathcal{T}_+ \mathcal{T}_- + \mathcal{T}_- \mathcal{T}_+ \right) \,.
\end{equation}

The generator $\Delta$ and the negative grade generators defined by it   remain unchanged
in going over to the deformed minreps.

%%%%%%%%%%%%%%%%%%%%
%%%%%%%%%%%%%%%%%%%%
The grade +1 generators are then given  by  the commutators :
\begin{equation}
\begin{split}
\widetilde{Q}
= i \commute{\,Q}{K_+}
& \qquad \qquad \qquad \qquad
\widetilde{Q}^\dagger
= \left( \widetilde{Q} \right)^\dag
= i \commute{\,Q^\dagger}{K_+}
\\
\widetilde{S}
= i \commute{\,S}{K_+}
& \qquad \qquad \qquad \qquad
\widetilde{S}^\dagger
= \left( \widetilde{S} \right)^\dag
= i \commute{\,S^\dagger}{K_+}
\end{split}
\end{equation}
Thus we make  an ansatz for grade +2 generator $K_+$  of the form 
\be
K_+ = \frac{1}{2}p^2 +\frac{1}{x^2}\left(c_1 \mathcal{T}^2+c_2 S^2+c_3 A^2 +c_4 \right) 
\ee
where  $c_1,..c_4$ are some constant parameters. 
Using the closure of the algebra, we determine these four unknown constants in terms of $\lambda$ and obtain:
\be
K_+ = \frac{1}{2}p^2+\frac{1}{4x^2}\left(8\lambda \mathcal{T}^2+\frac{8}{3}(\lambda-1)A^2+\frac{3}{2}+8\lambda(\lambda-1)S^2+2\lambda(\lambda-1)\right)
\ee
The +1 grade generators then take the form
\bea
\widetilde{Q} &=& -p\alpha +\frac{2i}{x}\left[\lambda\left\{\left(\mathcal{T}_0+\frac{3}{4}\right)\alpha +\mathcal{T}_-\beta \right\}-\frac{\lambda-1}{3}\left\{\left(A_0-\frac{3}{4}\right)\alpha -2A_-\beta^\dagger \right\}\right] \nn
\widetilde{Q}^\dagger &=& -p\alpha^\dagger-\frac{2i}{x}\left[\lambda\left\{\left(\mathcal{T}_0-\frac{3}{4}\right)\alpha^\dagger +\mathcal{T}_+\beta^\dagger \right\}-\frac{\lambda-1}{3}\left\{\left(A_0+\frac{3}{4}\right)\alpha^\dagger-2A_+\beta \right\}\right] \nn
\widetilde{S} &=& -p\beta -\frac{2i}{x}\left[\lambda\left\{\left(\mathcal{T}_0-\frac{3}{4}\right)\beta -\mathcal{T}_+\alpha \right\}-\frac{\lambda-1}{3}\left\{\left(A_0+\frac{3}{4}\right)\beta -A_-\alpha^\dagger \right\}\right] \nn
\widetilde{S}^\dagger &=& -p\beta^\dagger +\frac{2i}{x}\left[\lambda\left\{\left(\mathcal{T}_0+\frac{3}{4}\right)\beta^\dagger -\mathcal{T}_-\alpha^\dagger \right\}-\frac{\lambda-1}{3}\left\{\left(A_0-\frac{3}{4}\right)\beta^\dagger -A_-\alpha \right\}\right] \nn
\eea

The  anti-commutators of grade +1 and grade -1 generators close into grade zero subalgebra 
\bea
\anticommute{Q }{\widetilde{Q}^\dagger} &=& -\Delta -2i\lambda \mathcal{T}_0+i(\lambda+1)A_0 \nn 
\anticommute{Q^\dagger}{\widetilde{Q} } &=& -\Delta +2i\lambda \mathcal{T}_0-i(\lambda+1)A_0 \nn
\anticommute{S}{\widetilde{S}^\dagger} &=& -\Delta +2i\lambda \mathcal{T}_0+i(\lambda+1)A_0 \nn
\anticommute{S^\dagger}{\widetilde{S}} &=& -\Delta -2i\lambda \mathcal{T}_0-i(\lambda+1)A_0 
\eea
\be
\anticommute{Q}{\widetilde{S} } = +2i(\lambda+1)A_- 
\qquad \qquad\qquad \qquad\anticommute{Q }{\widetilde{S}^\dagger} = -2i\lambda \mathcal{T}_- 
\ee
\be
\anticommute{Q^\dagger}{\widetilde{S}^\dagger} = -2i(\lambda+1)A_+ 
\qquad \qquad\qquad \qquad \anticommute{Q^\dagger}{\widetilde{S} } = +2i\lambda \mathcal{T}_+ 
\ee
\be
\anticommute{S}{\widetilde{Q}} = -2i(\lambda+1)A_-
\qquad \qquad\qquad \qquad \anticommute{S}{\widetilde{Q}^\dagger} =-2i\lambda \mathcal{T}_+
\ee
\be
\anticommute{S^\dagger}{\widetilde{Q}^\dagger} = +2i(\lambda+1)A_+
\qquad \qquad\qquad \qquad \anticommute{S^\dagger}{\widetilde{Q}} = +2i\lambda \mathcal{T}_-
\ee
\be
\anticommute{Q}{\widetilde{Q}} = \anticommute{Q^\dagger}{\widetilde{Q}^\dagger} =\anticommute{S}{\widetilde{S}}=\anticommute{S^\dagger}{\widetilde{S}^\dagger}=0
\ee
\begin{equation}
\begin{aligned}
\commute{\widetilde{Q}}{K_-} &= i \, Q
\\
\commute{\widetilde{S}}{K_-} &= i \, S
\end{aligned}
\qquad \qquad
\begin{aligned}
\commute{\widetilde{Q}^\dagger}{K_-} &= i \, Q^\dagger
\\
\commute{\widetilde{S}^\dagger}{K_-} &= i \, S^\dagger
\end{aligned}
\end{equation}
\begin{center}
%\line(1,0){250}
\end{center}
The quadratic Casimir of $\mathfrak{su}(1,1)_K$ generated by $K_{\pm 2}$ and $\Delta$ is 
\bea
\mathcal{C}_2 \left[ \mathfrak{su}(1,1)_K \right]
&=& \mathcal{K}^2 = \frac{1}{2} (K_+ K_- + K_- K_+) - \frac{1}{4} \Delta^2 \nn
              &=&  \lambda \mathcal{T}^2+\frac{\lambda-1}{3}A^2+\lambda(\lambda-1)S^2+\frac{\lambda(\lambda-1)}{4}
\eea

There exists a one parameter family of quadratic Casimir elements $C_2(\mu)$  that commute with all the generators  of $D(2,1; \lambda) $. 
\bea \label{C2}
\mathcal{C}_2 (\mu)  &=&\frac{\mu}{4}  \mathcal{K}^2  -\frac{ \lambda}{4} (\mu -8)  \mathcal{T}^2 - \frac{1}{12} \left(16 + 8 \lambda+ \mu (\lambda -1))\right)A^2 +\frac{i}{4}\mathcal{F}(Q,S) \nn
&=& \frac{\lambda}{4} \left( 8 +\mu(\lambda-1)\right)\left(S^2+\frac{1}{4}\right)
\eea
where
 \bea
\mathcal{F}(Q,S) &=& \commute{Q}{\widetilde{Q}^\dagger}+\commute{Q^\dagger}{\widetilde{Q}}+\commute{S}{\widetilde{S}^\dagger}+\commute{S^\dagger}{\widetilde{S}}
\eea
is the contribution from the odd generators.
Since the eigenvalues of the quadratic Casimir depends on the eigenvalues $s(s+1)$ of the Casimir operator $S^2$ of $SU(2)_S$ the corresponding deformed unitary supermultiplets will be uniquely labelled by spin $s$ of $SU(2)_S$ . 
%%%%%%%%%%%%%%%%%%%%%%%%%%%%%%%%
\section{$SU(2)$ deformed minimal unitary representations as positive energy unitary supermultiplets of $D(2,1;\lambda)$ }
\subsection{ Compact 3-grading }
As  shown in above the Lie superalgebra $D(2,1;\lambda)$ admits a 5-graded decomposition
defined by the  generator $\Delta$ :
\begin{equation}
\begin{split}
D(2,1;\lambda)
&= \mathfrak{g}^{(-2)} \oplus
   \mathfrak{g}^{(-1)} \oplus
   \left[ \mathfrak{su}(2)_{\mathcal{T}} \oplus
	    \mathfrak{su}(2)_A \oplus
          \mathfrak{so}(1,1)_\Delta
   \right] \oplus
   \mathfrak{g}^{(+1)} \oplus
   \mathfrak{g}^{(+2)}
\\
&= K_-
   \oplus
   \left[Q \,,\, Q^\dagger \,,\, S \,,\, S^\dagger \right]
\oplus
   \left[ A_{\pm,0} \,,\, \mathcal{T}_{\pm,0} \,,\, \Delta
   \right]
   \oplus
   \left[\widetilde{Q} \,,\, \widetilde{Q}^\dagger \,,\,
          \widetilde{S} \,,\, \widetilde{S}^\dagger
   \right]
   \oplus
   K_+  \nonumber
\end{split}
\end{equation}

The Lie superalgebra $D(2,1;\lambda)$ can also be a given a  3-graded decomposition with respect to its compact subsuperalgebra $\mathfrak{osp}(2|2)\oplus \mathfrak{u}(1) = \mathfrak{su}(2|1) \oplus \mathfrak{u}(1)$ , which we shall refer to   as compact  3-grading :
\be
D(2,1;\lambda)
= \mathfrak{C}^- \oplus \mathfrak{C}^0 \oplus \mathfrak{C}^+
\ee
\be
D(2,1;\lambda) = \left( A_-, B_-, \mathfrak{Q}_- ,\mathfrak{S}_- \right) \oplus 
\left( \mathcal{T}_{\pm,0} ,\mathcal{J} , \mathcal{H},\mathfrak{Q}_0 ,\mathfrak{S}_0 ,\mathfrak{Q}_0^\dagger ,\mathfrak{S}_0^\dagger \right) \oplus \left( A_+, B_+, \mathfrak{Q}_+ ,\mathfrak{S}_+ \right) 
\ee

The generators belonging to grade -1 subspace $\mathfrak{C}^-$ are  as follows:
\bea
 A_- &= &\beta \alpha  \\
B_-
  &=& \frac{i}{2} \left[ \Delta + i \left( K_+ - K_- \right) \right]  \\ \nonumber
 &=& \frac{1}{4} \left( x + i p \right)^2
    - \frac{1}{ \, x^2} \left(  \, L^2 + \frac{3}{16} \right) 
     \\
\mathfrak{Q}_-  &=& \frac{1}{2}(Q-i\widetilde{Q}) \\
&=& \frac{1}{2}(x+ip)\alpha +\frac{1}{x}\left[
      \left(\frac{2\lambda+1}{3}\left\{ G_0 +\frac{3}{4}\right\} +\lambda S_0 \right) \alpha
      + \left\{\frac{2\lambda+1}{3} G_- + \lambda S_-\right\} \beta \right] \nn 
&=&\frac{1}{2}(x+ip)\alpha +\frac{1}{x}\left[(2\lambda+1)\left(\frac{1}{4}-\frac{\beta^\dagger \beta}{2}\right)\alpha + \lambda\left(\frac{(a^ma_m-b^mb_m)\alpha}{2}+b^ma_m\beta \right)\right] \nn
\mathfrak{S}_- &=& \frac{1}{2}(S-i\widetilde{S}) \\
&=& \frac{1}{2}(x+ip)\beta -\frac{1}{x}\left[
      \left(\frac{2\lambda+1}{3}\left\{ G_0 -\frac{3}{4}\right\} +\lambda S_0 \right) \beta
      - \left\{\frac{2\lambda+1}{3} G_+ + \lambda S_+\right\} \alpha \right] \nn
&=&\frac{1}{2}(x+ip)\beta -\frac{1}{x}\left[(2\lambda+1)\left(\frac{\alpha^\dagger \alpha}{2}-\frac{1}{4}\right)\beta + \lambda\left(\frac{(a^ma_m-b^mb_m)\beta}{2}-a^mb_m\alpha\right)\right] \nonumber
\eea
where 
\be 
 L^2=\lambda \mathcal{T}^2+\frac{1}{3}(\lambda-1)A^2+\lambda(\lambda-1)S^2+\frac{1}{4}\lambda (\lambda-1)
 \ee
 The grade +1  generators in $\mathfrak{C}^+$ are obtained by Hermitian conjugation of grade $-1$ generators:
\bea
A_+ &=& \alpha^\dagger \beta^\dagger \\
B_+ &=& - \frac{i}{2} \left[ \Delta - i \left( K_+ - K_- \right) \right] \\
 &=& \frac{1}{4} \left( x - i p \right)^2
    - \frac{1}{ \, x^2} \left(  \, L^2 + \frac{3}{16} \right)
\nn
\mathfrak{Q}_+ &=& (\mathfrak{Q}_-)^\dagger = \frac{1}{2}(Q^\dagger+i\widetilde{Q}^\dagger) \\
&=& \frac{1}{2}(x-ip)\alpha^\dagger +\frac{1}{x}\left[
      \left(\frac{2\lambda+1}{3}\left\{ G_0 -\frac{3}{4}\right\} +\lambda S_0 \right) \alpha^\dagger
      + \left\{\frac{2\lambda+1}{3}G_+ +\lambda S_+\right\} \beta^\dagger \right] \nn
&=&\frac{1}{2}(x-ip)\alpha^\dagger +\frac{1}{x}\left[(2\lambda+1)\left(\frac{1}{4}-\frac{\beta^\dagger \beta}{2}\right)\alpha^\dagger + \lambda\left(\frac{(a^ma_m-b^mb_m)\alpha^\dagger}{2}+a^mb_m\beta^\dagger\right)\right] \nn
\mathfrak{S}_+ &=& (\mathfrak{S}_-)^\dagger = \frac{1}{2}(S^\dagger+i\widetilde{S}^\dagger) \\
&=& \frac{1}{2}(x-ip)\beta^\dagger -\frac{1}{x}\left[
      \left(\frac{2\lambda+1}{3}\left\{ G_0 +\frac{3}{4}\right\} +\lambda S_0 \right) \beta^\dagger
      - \left\{\frac{2\lambda+1}{3} G_- + \lambda S_-\right\} \alpha^\dagger \right] \nn
&=&\frac{1}{2}(x-ip)\beta^\dagger -\frac{1}{x}\left[(2\lambda+1)\left(\frac{\alpha^\dagger \alpha}{2}-\frac{1}{4}\right)\beta^\dagger + \lambda\left(\frac{(a^ma_m-b^mb_m)\beta^\dagger}{2}-b^ma_m\alpha^\dagger\right)\right] \nonumber \eea
The grade 0 fermionic generators in $ \mathfrak{C}^0$ are given by
\bea
\mathfrak{Q}_0 &=& \frac{1}{2}(Q+i\widetilde{Q}) \\
&=& \frac{1}{2}(x-ip)\alpha -\frac{1}{x}\left[
      \left(\frac{2\lambda+1}{3}\left\{ G_0 +\frac{3}{4}\right\} + \lambda S_0 \right) \alpha
      + \left\{ \frac{2\lambda+1}{3}G_- + \lambda S_-\right\} \beta \right] \nn
&=&\frac{1}{2}(x-ip)\alpha -\frac{1}{x}\left[(2\lambda+1)\left(\frac{1}{4}-\frac{\beta^\dagger \beta}{2}\right)\alpha + \lambda\left(\frac{(a^ma_m-b^mb_m)\alpha}{2}+b^ma_m\beta\right)\right] \nonumber
\eea
\bea
\mathfrak{S}_0&=& \frac{1}{2}(S+i\widetilde{S}) \\
&=& \frac{1}{2}(x-ip)\beta +\frac{1}{x}\left[
      \left(\frac{2\lambda+1}{3}\left\{G_0 -\frac{3}{4}\right\} + \lambda S_0 \right) \beta
      - \left\{\frac{2\lambda+1}{3} G_+ + \lambda S_+\right\} \alpha \right] \nn
&=&\frac{1}{2}(x-ip)\beta +\frac{1}{x}\left[(2\lambda+1)\left(\frac{\alpha^\dagger \alpha}{2}-\frac{1}{4}\right)\beta + \lambda\left(\frac{(a^ma_m-b^mb_m)\beta}{2}-a^mb_m\alpha\right)\right] \nonumber
\eea
\bea
\mathfrak{Q}_0^\dagger &=& \frac{1}{2}(Q^\dagger-i\widetilde{Q}^\dagger) \\
&=& \frac{1}{2}(x+ip)\alpha^\dagger -\frac{1}{x}\left[
      \left(\frac{2\lambda+1}{3}\left\{ G_0 -\frac{3}{4}\right\} + \lambda S_0 \right) \alpha^\dagger
      + \left\{\frac{2\lambda+1}{3}G_+ + \lambda S_+\right\} \beta^\dagger \right] \nn
&=&\frac{1}{2}(x+ip)\alpha^\dagger -\frac{1}{x}\left[(2\lambda+1)\left(\frac{1}{4} - \frac{\beta^\dagger \beta}{2}\right)\alpha^\dagger + \lambda\left(\frac{(a^ma_m-b^mb_m)\alpha^\dagger}{2}+a^mb_m\beta^\dagger\right)\right] \nonumber
\eea
\bea
\mathfrak{S}_0^\dagger &=& \frac{1}{2}(S^\dagger-i\widetilde{S}^\dagger) \\
&=& \frac{1}{2}(x+ip)\beta^\dagger +\frac{1}{x}\left[
      \left(\frac{2\lambda+1}{3}\left\{ G_0 +\frac{3}{4}\right\} + \lambda S_0 \right) \beta^\dagger
      - \left\{\frac{2\lambda+1}{3} G_- + \lambda S_-\right\} \alpha^\dagger \right] \nn
&=&\frac{1}{2}(x+ip)\beta^\dagger +\frac{1}{x}\left[(2\lambda+1)\left(\frac{\alpha^\dagger \alpha}{2}-\frac{1}{4}\right)\beta^\dagger + \lambda\left(\frac{(a^ma_m-b^mb_m)\beta^\dagger}{2}-b^ma_m\alpha^\dagger\right)\right]
 \nonumber
\eea
The  grade zero odd generators together with the $SU(2)_{\mathcal{T}}$ generators  $\mathcal{T}_+, \mathcal{T}_- , \mathcal{T}_0$ and $U(1)$ generator 
\be 
\mathcal{J} = (\lambda+1)A_0 +\frac{1}{2} \left( K_+ + K_- \right)
\ee
generate the sub-supergroup $ SU(2|1)$ .
 They satisfy the anticommutation relations
\be
\begin{array}{cclccl}
\anticommute{\mfQ_0}{\mfQd_0} &=& -\lambda \mathcal{T}_0+\mcJ & \anticommute{\mfQ_0}{\mfSd_0} &=& -\lambda \mathcal{T}_- \\
\anticommute{\mfS_0}{\mfSd_0} &=& +\lambda \mathcal{T}_0+\mcJ & \anticommute{\mfS_0}{\mfQd_0} &=& -\lambda \mathcal{T}_+ \\
\commute{\mathcal{T}_0}{\mfQ_0} &=& -\half\mfQ_0 & \commute{\mathcal{T}_0}{\mfS_0} &=&+ \half\mfS_0\\
\commute{\mathcal{T}_0}{\mfQd_0} &=& +\half\mfQd_0 & \commute{\mathcal{T}_0}{\mfSd_0} &=& -\half\mfSd_0
\\
\commute{\mcJ}{\mfQ_0} &=& -\frac{\lambda}{2}\mfQ_0 & \commute{\mcJ}{\mfS_0} &=& -\frac{\lambda}{2}\mfS_0\\
\commute{\mcJ}{\mfQd_0} &=& +\frac{\lambda}{2}\mfQd_0 & \commute{\mcJ}{\mfSd_0} &=& +\frac{\lambda}{2}\mfSd_0\\
\commute{\mathcal{T}_+}{\mfQ_0} &=& -\mfS_0 & \commute{\mathcal{T}_-}{\mfS_0} &=&-\mfQ_0\\
\commute{\mathcal{T}_+}{\mfQd_0} &=& +\mfSd_0 & \commute{\mathcal{T}_-}{\mfSd_0} &=& +\mfQd_0
\end{array}
\ee
The anticommutation relations of  grade zero fermionic generators with grade $\pm 1$ generators in the compact 3-grading are as follows
\be
\begin{array}{cclccl}
\anticommute{\mfQ_0}{\mfQ_+} &=& 2B_+ & \anticommute{\mfS_0}{\mfQ_+} &=& 0 \\
\anticommute{\mfQd_0}{\mfQ_+} &=& 0 & \anticommute{\mfSd_0}{\mfQ_+} &=& 2(\lambda+1)A_+ \\
\anticommute{\mfQ_0}{\mfS_+} &=& 0 & \anticommute{\mfS_0}{\mfS_+} &=& 2B_+ \\
\anticommute{\mfQd_0}{\mfS_+} &=& -2(\lambda+1)A_+ & \anticommute{\mfSd_0}{\mfS_+} &=& 0  \\
\anticommute{\mfQ_0}{\mfQ_-} &=& 0 & \anticommute{\mfS_0}{\mfQ_-} &=& 2(\lambda+1)A_- \\
\anticommute{\mfQd_0}{\mfQ_-} &=& 2B_- & \anticommute{\mfSd_0}{\mfQ_-} &=& 0  \\
\anticommute{\mfQ_0}{\mfS_-} &=& -2(\lambda+1)A_- & \anticommute{\mfS_0}{\mfS_-} &=& 0 \\
\anticommute{\mfQd_0}{\mfS_-} &=& 0 & \anticommute{\mfSd_0}{\mfS_-} &=& 2B_-  \\
\end{array}
\ee
The generator $\mathcal{H}$ that determines the compact three grading is given by 
\bea
\mathcal{H}
&=& \frac{1}{2} \left( K_+ + K_- \right)
   + A_0 \nn
&=& B_0 + \frac{1}{2}\left(N_\alpha+N_\beta-1\right) 
\eea
where 
\bea
B_0 &=& \frac{1}{4}(p^2+x^2)+\frac{1}{x^2}\left(L^2+\frac{3}{16}\right)\\
 &=& \frac{1}{4}(p^2+x^2)+\frac{1}{x^2}\left(\lambda \mathcal{T}^2+\frac{1}{3}(\lambda-1)A^2+\lambda(\lambda-1)S^2+\frac{1}{4}\lambda (\lambda-1)+\frac{3}{16}\right) \nonumber
\eea

\subsection{ Unitary supermultiplets of $D(2,1;\lambda)$ }

The generators  $B_-$ and $B_+$ defined  above close into  $B_0$ under commutation and  generate the distinguished  $\mathfrak{su}(1,1)_K$
subalgebra
\begin{equation}
\commute{B_-}{B_+} = 2 \, B_0
\qquad \qquad
\commute{B_0}{B_+} = + \, B_+
\qquad \qquad
\commute{B_0}{B_-} = - \, B_- \,.
\end{equation}
%%%%%%%%%%TBC%%%%%%%%%%
The generator   $B_0$  can be interpreted as $\frac{1}{2}$  the Hamiltonian , $H_{Conf}$ , 
 of conformal quantum mechanics \cite{deAlfaro:1976je} or of a singular oscillator \cite{MR858831}  
 \be
 H_{Conf} = 2 B_0=  \frac{1}{2} \left( x^2 + p^2 \right) + \frac{g^2}{x^2} 
 \ee
 with $ g^2 =(2L^2 +\frac{3}{8} )$ playing the role of  coupling constant.  
A unitary lowest weight ( positive energy) irreducible representation of $SU(1,1)_K$ is uniquely determined by the state $|\psi^\alpha_0 \rangle$ with the lowest eigenvalue of $B_0$  that is  annihilated by $B_-$:
\be
B_ - |\psi^\omega_0 \rangle =0
\ee
 In the coordinate ($x$) representation its wave function is  given by
\be
\psi_0^\omega (x) = C_0 \, x^{\omega} \,e^{-x^2/2}
\ee
where $C_0$ is  the normalization constant, $\omega$ is given by 
\be \label{omegasol}
\omega = \frac{1}{2} + \left( \frac{1}{4} + 2 \hat{g}^2 \right)^{1/2} 
\ee
and $\hat{g}^2$ is the eigenvalue of   $(2L^2 +\frac{3}{8} )$ 
\be (2L^2 +\frac{3}{8} ) |\psi_0^\omega \rangle = \hat{g}^2 |\psi_0^\omega \rangle \ee

We shall denote the functions obtained by the repeated action of differential operators $B_+$ on $\psi_0^\omega(x)$ in the coordinate representation as $\psi_n^\omega (x) $ and the corresponding states in the Hilbert space as $|\psi_n^\omega\rangle $ :
\be 
\psi_n^\omega(x) = c_n (B_+)^n \psi_0^\omega(x)
\ee
where the normalization constant is given as
\be
c_n = \frac{(-1)}{2^n} \frac{\sqrt{\Gamma(\omega + 1/2)}}{\sqrt{n! \,  \Gamma(n+\omega +1/2)} }
\ee
The wave functions $ \psi_n^\omega(x)$ can be written as
\be
\psi_n^\omega(x) = \sqrt{\frac{2 (n!) } {\Gamma( n+ \omega +1/2)}} \, L_n^{(\omega -1/2)}(x^2)  x^\omega e^{- x^2/2} 
\ee
where $L_n^{(\omega -1/2)}(x^2) $ is the generalized Laguerre polynomial.

Irreducible unitary lowest weight representations of $D(2,1;\lambda)$  are uniquely labelled by a set of states , which we shall simply denote as $ \{|\Omega \rangle \}$ , that transform irreducibly under the grade zero compact subsupergroup $OSp(2/2) \times U(1)$ and  are annihilated by the grade -1 generators $ B_-, A_-, \mathfrak{Q}_-$ and $\mathfrak{S}_-$ in $\mathfrak{C}^-$.  
In a unitary lowest weight ( positive energy) representation of $D(2,1;\lambda)$ the spectrum of $\mathcal{H}$ is bounded from below. We shall refer to $\mathcal{H}$ as the (total) Hamiltonian and its eigenvalues as total {\it energy}. 
Each state in the set $ {|\Omega \rangle }$ is a lowest (conformal) energy  state  of  a positive energy irrep of $SU(1,1)_K$, since they are all annihilated by $B_-$.  The conditions 
\bea \label{grade-1}
B_- |\Omega\rangle  =0 \nn 
A_- |\Omega\rangle  =0 \nn 
\mathfrak{Q}_- |\Omega\rangle =0  \nn
\mathfrak{S}_- | \Omega \rangle =0 
\eea
imply that the states $|\Omega \rangle $ must be linear combinations of the  tensor product states of 
the form
\be
|F\rangle \times |B\rangle \times |\psi^\omega_0 \rangle  \label{lowestF} \
\ee
where the state $|F\rangle $ in (\ref{lowestF}) is either the fermionic Fock vacuum
\be 
 | 0\rangle_F = |m_t= 0;m_a=-1/2\rangle_F = \ket{0,\dowa}
\ee
or one of the following  $SU(2)_T$ doublet of states:
\bea
  \alpha^\dagger | 0\rangle_F \equiv |m_t=1/2;m_a=0\rangle_F=\ket{\upa,0}, \\
   \beta^\dagger | 0\rangle_F \equiv |m_t=-1/2;m_a=0\rangle_F=\ket{\dowa,0}
   \eea
and 
the state $|B\rangle$ in (\ref{lowestF})  is any one of the states 

\be
a^{m_1} \cdots a^{m_k} b^{m_{k+1}} \cdots b^{m_{2s}} |0\rangle_B \label{SU2S}
\ee
where   $|0\rangle_B$ is the bosonic Fock vacuum annihilated by the bosonic oscillators   $a_m $ and $b_m$ $ ( m =1,2, \cdots , P)$.  For fixed $n$ the states of the form (\ref{SU2S}) transform in the spin  $ s $ representation  of $SU(2)_S $. They also form representations of  $SO^*(2P)$ generated by the bilinears
$
U^m_{~ n} = a^m a_n + b^m b_n 
$ , $(a_m b_n -a_n b_m)$ and $(a^m b^n -a^n b^m)$
and which commutes with   $D(2,1;\lambda)$. Therefore as far as the $SU(2)$ deformations of the minrep of $D(2,1;\lambda)$ are concerned we can restrict our analysis to $P=1$.
Then for $P=1$  we simply have 
\[ S_+ =  a^\dagger b \]
\[ S_-=b^\dagger a \]
and \[ S_0 = \frac{1}{2}( a^\dagger a - b^\dagger b) \]
Then the states $|B\rangle$ belong to the set
\be
| s, m_s\rangle_B \equiv ( a^\dagger )^k \, (b^\dagger)^{2s-k} |0\rangle_B 
\ee
where $ m_s=k-s  $ ( $k=0,1, \cdots , 2s$ ) and  transform in spin $s$ representation of $SU(2)_S$: 
\be
S^2 | s, m_s \rangle_B = s(s+1) | s, m_s \rangle_B
\ee
\be
S_0 | s, m_s \rangle_B = m_s| s, m_s \rangle_B 
\ee
The action of raising operator $S_+=a^\dagger b$ and lowering operator $S_-=b^\dagger a$ on this state is then given as
\bea
S_+\ket{s,m_s=k-s}_B \eq \sqrt{(k+1)(2s-k)}\ket{s,k-s+1}_B \\
S_- \ket{s,m_s=k-s}_B \eq \sqrt{k(2s-k+1)}\ket{s,k-s-1}_B
\eea
The eigenvalues $(1/4 + 2\hat{g}^2)$ of $(4L^2+1) $ on the above states determine the values of $\omega$ labeling the  eigenstates $|\psi^\omega_0 \rangle $ of $B_0$ annihilated by $B_-$  :
\be 
( 4L^2 + 1) |m_t=0;m_a= -1/2 \rangle_F \times  |s, m_s\rangle_B = \lambda ^2 (2s+1)^2 |0;-1/2\rangle_F \times   |s,m_s\rangle_B  \label{mgzero}
\ee
\be 
( 4L^2 + 1) |m_t= \pm 1/2;m_a=0 \rangle_F \times  |s, m_s\rangle_B = [ \lambda (2s+1)+1 ]^2 |\pm 1/2 ;0\rangle_F \times   |s,m_s\rangle_B \label{mghalf}
\ee
The $SU(2)$ subalgebra of $SU(1|2)$  is the diagonal subalgebra $ SU(2)_{\mathcal{T}}$ of $SU(2)_S$ and $SU(2)_T$. Therefore we shall work in a basis where $\mathcal{T}^2$ and $\mathcal{T}_0$ are diagonalized and   denote the simultaneous eigenstates of $B_0$, $\mathcal{T}^2 , \mathcal{T}_0$ , $A^2$ and $A_0$ as\footnote{We introduce this notation for the states because the tensor product states of the form $\ket{F}\times\ket{B}\times\ket{\psi_0^\omega}$ are not always definite eigenstates of $\mathcal{T}^2 $and $ \mathcal{T}_0$ but it is easier to understand the structure of supermultiplets in terms of these tensor product states. Thus we will use both notations for states.}
\be
|\omega; \mathfrak{t}, m_{\mft}; a, m_a\rangle
\ee
where $\omega$ is the eigenvalue of $B_0$.

The set of states $\ket{\Omega}$ must be linear combinations of the tensor product states of the form $\ket{F}\times\ket{B}\times\ket{\psi_o^\omega}$ where the state $\ket{F}$ could be either $\ket{0}_F=\ket{0,\dowa}_F$ or one of the $SU(2)_T$ doublet of states $\ket{\upa,0}_F$ or $\ket{\dowa,0}_F$. We will now study the unitary representations for these two cases.
%%%%%%%%%%%%%%%%%%%%%%%%%%%%%%%%%%%%%%%%%%%%%%%%%%%%%%%%%%%%%%%%%%%%%% 
\\

\subsubsection{\bf $\ket{F}=\ket{0}_F$}

For states with $\mft=0$, we can use equation (\ref{omegasol}) to write
\be \omega = \frac{1}{2} \pm \lambda ( 2s+1) \ee
where the sign of the square root is determined by the sign of $\lambda$ and the range of $\lambda$ is determined by the square integrability of the states and the positivity of $\hat{g}^2$. This leads to the following restriction on $\lambda$

\be
|\lambda| >\frac{1}{2(2s+1)}
\ee

Let us first consider the case $s=0$ and with the positive square root taken in the above equations. Then the lowest energy state
$\tensor{0,\dowa}{0}{\lambda+1/2}{0}$ , annihilated by all the generators in $\mathfrak{C}^{-1}$ , is a singlet of the grade zero super algebra $SU(1|2)$ since

\bea
\mathfrak{Q}_0 \tensor{0,\dowa}{0}{\lambda+1/2}{0} &=& 0 \nn
\mathfrak{S}_0 \tensor{0,\dowa}{0}{\lambda+1/2}{0} &=& 0 \nn
\mathfrak{Q}_0^\dagger\tensor{0,\dowa}{0}{\lambda+1/2}{0} &=& 0 \nn
\mathfrak{S}_0^\dagger\tensor{0,\dowa}{0}{\lambda+1/2}{0} &=& 0
\eea
States generated by action of grade +1 generators $\mathfrak{C}^+$ on this lowest weight state $\ket{\psi_0^{(\lambda+1/2)}}$ 
\bea
B^+\tensor{0,\dowa}{0}{\lambda+1/2}{0} &=& \tensor{0,\dowa}{0}{\lambda+1/2}{1}   \nn
A^+\tensor{0,\dowa}{0}{\lambda+1/2}{0} &=& \tensor{0,\upa}{0}{\lambda+1/2}{0}   \nn
\mathfrak{Q}_+\tensor{0,\dowa}{0}{\lambda+1/2}{0} &=& \tensor{\upa,0}{0}{\lambda+3/2}{0}  \nn 
\mathfrak{S}_+\tensor{0,\dowa}{0}{\lambda+1/2}{0}&=&  \tensor{\dowa,0}{0}{\lambda+3/2}{0}
\eea
form a supermultiplet  transforming in the representation with  super tableau 
 $\soneonebox$ of $SU(2|1)$ .
The states $\mathfrak{Q}_+\tensor{0,\dowa}{0}{\lambda+1/2}{0} $ and $\mathfrak{S}_+\tensor{0,\dowa}{0}{\lambda+1/2}{0}$ are both lowest weight vectors of $SU(1,1)_K$ transforming as a doublet of $SU(2)_T$ . The state $A^+\tensor{0,\dowa}{0}{\lambda+1/2}{0}$ is a lowest weight vector of $SU(1,1)_K$  and together with $\tensor{0,\dowa}{0}{\lambda+1/2}{0}$ form a doublet of $SU(2)_A$.
The commutator of two susy generators in  $\mathfrak{C}^+$ satisfies 
\[ {[} \mathfrak{Q}_+, \mathfrak{S}_+ {]} \tensor{0,\dowa}{0}{\lambda+1/2}{0} \,\,  \propto \,\,\alpha^\dagger\beta^\dagger B^+ \tensor{0}{0}{\lambda+1/2}{0}  \]
Hence one does not generate any new lowest weight vectors of $SU(1,1)_K$ by further actions of grade $+1$ supersymmetry generators.

Every positive energy unitary representation of the conformal group $SO(d,2)$ corresponds to a conformal field in $d$ dimensional Minkowskian spacetime. The eigenvalues of the $SO(2)$ generator determine the conformal dimension of the field.  In one dimension  the positive energy unitary representations of $SO(2,1)$ are identified with conformal wave functions. We shall denote the conformal wave function associated with a positive energy unitary representation of $SO(2,1)$ with lowest weight vector $\ket{\psi_0^{(\omega)}}$ as $\Psi^{\omega}(x)$. The conformal wave functions transforming in the 
$(\mft,a)$ representation of $SU(2)_{\mathcal{T}}\times SU(2)_A$ will then be denoted as 
\[ \Psi_{(\mft,a)}^{\omega}(x) \]
Thus the unitary  supermultiplet of $D(2,1;\lambda)$ with the lowest weight vector $\tensor{0}{0}{\lambda+1/2}{0}$ decomposes as:
\be
 \Psi^{(\lambda + 1)/2}_{(0,1/2)} \oplus \Psi^{(\lambda + 2)/2}_{(1/2,0)}
\ee
This is simply the minimal unitary  supermultiplet of $D(2,1;\lambda)$ and for $\lambda=-1/2$ coincides with  the singleton supermultiplet of $OSp(4|2,\mathbb{R})= D(2,1;-1/2)$. 
%%%%%%%%%%%TBC

Next we consider the representations for the case $s\neq 0$ with $\lambda>0$. In this case the lowest energy states 
\be
\tensor{0,\dowa}{s,m_s=(k-s)}{\omega}{0},
\ee
where $\omega = 1/2+\lambda(2s+1)$, 
and $k=0,\ldots,2s$, are not annihilated by all supersymmetry generators of $SU(2|1)$:

\bea \label{grade0}
\mathfrak{Q}_0 \tensor{0,\dowa}{s,k-s}{\omega}{0} &=& 0 \nn
\mathfrak{S}_0 \tensor{0,\dowa}{s,k-s}{\omega}{0} &=& 0 \nn
\mathfrak{Q}_0^\dagger \tensor{0,\dowa}{s,k-s}{\omega}{0} &=& \lambda(2s-k)\tensor{\upa,0}{s,k-s}{\omega-1}{0} \nn&&- \lambda\sqrt{(k+1)(2s-k)}\tensor{\dowa,0}{s,k-s+1}{\omega-1}{0} \nn
\mathfrak{S}_0^\dagger \tensor{0,\dowa}{s,k-s}{\omega}{0} &=& \lambda k\tensor{\dowa,0}{s,k-s}{\omega-1}{0} \nn
&&- \lambda\sqrt{k(2s-k+1)}\tensor{\upa,0}{s,k-s-1}{\omega-1}{0}\nn
\commute{\mathfrak{Q}_+}{\mathfrak{S}_+} \tensor{0,\dowa}{s,k-s}{\omega}{0} \eq 2s \tensor{0,\upa}{s,k-s}{\omega-2}{0}
\eea

Since the supersymmetry generators $\mathfrak{Q}_0^\dagger $ and $\mathfrak{S}_0^\dagger $ transform in the spin 1/2  representation of $SU(2)_{\mathcal{T}}$  acting on the states with spin $\mft=s$  one would expect to obtain states with both $\mft=s\pm1/2$.  However setting $k=2s$ in the above formulas we find
\bea
\mathfrak{Q}_0^\dagger \tensor{0,\dowa}{s,s}{\omega}{0} &=& 0 \nn
\mathfrak{S}_0^\dagger \tensor{0,\dowa}{s,s}{\omega}{0} &=& 2 \lambda s \tensor{\dowa,0}{s,s}{\omega-1}{0} \nn
&&- \lambda\sqrt{2s}\tensor{\upa,0}{s,s-1}{\omega-1}{0}\nn
\eea
which implies that we only get states with $\mft=s-1/2$. 
Therefore the lowest energy supermultiplets of $SU(2|1)$ that uniquely determine the deformed minimal unitary supermultiplets of $D(2,1;\lambda)$ transform in the representation with the super tableau $\underbrace{\sgenrowbox}_{2s}$ which decomposes under the even subgroup 
$SU(2)_{\mathcal{T}} \times U(1)_{\mathcal{J}}$ as
\bea
\underbrace{\sgenrowbox}_{2s} = (\underbrace{\genrowbox}_{2s}, 0) \oplus (\underbrace{\genrowbox}_{(2s-1)}, \frac{\lambda}{2} )
\eea

%%%%%%%%%%%%%%%%%%%TBC%%%%%%%%%%%%%%
%%%%%%%%%%%%%%%%%%%%%%%%%%%%%%%%
By acting  with  grade +1 generators of the compact 3-grading on these states with $\mft=s$ and $\mft=s-1/2$ to obtain states with $\mft=s\pm1/2$ and $\mft=s$ :

\bea
B^+\tensor{0,\dowa}{s,k-s}{\omega}{0} &=& \tensor{0,\dowa}{s,k-s}{\omega}{1}   \nn
A^+ \tensor{0,\dowa}{s,k-s}{\omega}{0} &=& \tensor{0,\upa}{s,k-s}{\omega}{0} \nn
\mathfrak{Q}_+\tensor{0,\dowa}{s,k-s}{\omega}{0} &=& \tensor{\upa,0}{s,k-s}{\omega+1}{0}\nn &&-\lambda (2s-k) \tensor{\upa,0}{s,k-s}{\omega-1}{0} \nn
&& +\lambda \sqrt{(k+1)(2s-k)}\tensor{\dowa,0}{s,k-s+1}{\omega-1}{0} \nn 
\mathfrak{S}_+\tensor{0,\dowa}{s,k-s}{\omega}{0} &=& \tensor{\dowa,0}{s,k-s}{\omega+1}{0} \nn 
&& -\lambda(2s- k) \tensor{\dowa,0}{s,k-s}{\omega-1}{0} \nn
&&+\lambda\sqrt{k(2s-k+1)}\tensor{\upa,0}{s,k-s-1}{\omega-1}{0} \nn
\eea
The commutator of two supersymmetry generators does not generate any new lowest weight vector of $SU(1,1)_K$:
\bea   
\commute{\mathfrak{Q}_+}{\mathfrak{S}_+} \tensor{0,\dowa}{s,k-s}{\omega}{0} &\propto& \tensor{0,\upa}{s,k-s}{\omega}{1}= \nn
&& \alpha^\dagger\beta^\dagger B^+ \tensor{0,\dowa}{s,k-s}{\omega}{0}
\eea 
Thus the complete supermultiplet is simply 
\be
\Psi^{p}_{(s-1/2,0)} \oplus \Psi^{p+1/2}_{(s,1/2)} \oplus \Psi^{p+1}_{(s+1/2,0)} 
\ee
where $p=\lambda(2s+1)/2$. We have summarized the deformed supermultiplets for lowest weight states with $\mft=s$ and $\lambda>1/(4s+2)$ in Table \ref{table-spositive}. 

\begin{table}[htp]
\caption{  \label{table-spositive} Decomposition of $SU(2)$  deformed minimal unitary  lowest energy supermultiplets of $D(2,1;\lambda)$  with respect to $SU(2)_{\mathcal{T}}\times SU(2)_A\times SU(1,1)_K$ . The conformal wavefunctions transforming in the $(\mft,a)$ representation of $SU(2)_{\mathcal{T}}\times SU(2)_A$ with conformal energy $\omega$ are denoted as $\Psi_{(\mft,a)}^{\mathfrak{\omega}}$.
The first column shows the super tableaux  of the lowest energy $SU(2|1)$ supermultiplet, the second column gives the eigenvalue of the $U(1)$ generator $\mathcal{H}$. The allowed range of $\lambda$ in this case is $\lambda>1/(4s+2)$.} 
\begin{center}
\begin{tabular}{|c|c|c|}
\hline
 $SU(2|1) l.w.v$ & $\mathcal{H}$ & $SU(1,1)_K\times SU(2)_\mathcal{T}\times SU(2)_A$\\
\hline
&& \\
 $1$ & $\lambda/2$ & $\Psi_{(0,1/2)}^{(\lambda+1)/2}\oplus\Psi_{(1/2,0)}^{(\lambda+2)/2}$ \\ 
&& \\
\hline
&& \\
 $\sonebox$ & $\lambda$ & $\Psi_{(0,0)}^{\lambda}\oplus\Psi_{(1/2,1/2)}^{\lambda+1/2}\oplus  \Psi_{(1,0)}^{\lambda+1}$ \\
&& \\
\hline
&& \\
 $\stwobox$ & $3\lambda/2$ & $\Psi_{(1/2,0)}^{3\lambda/2}\oplus \Psi_{(1,1/2)}^{(3\lambda+1)/2}\oplus \Psi_{(3/2,0)}^{(3\lambda/2+1}$ \\
&& \\
\hline
$\vdots$ & $\vdots$ & $\vdots$ \\
$\vdots$ & $\vdots$ & $\vdots$ \\
\hline
&&\\
 $\underbrace{\sgenrowbox}_{2s}$ & $(2s+1)\lambda/2$ & $\Psi^{p}_{(s-1/2,0)} \oplus \Psi^{p+1/2}_{(s,1/2)} \oplus \Psi^{p+1}_{(s+1/2,0)}$ \\
&& $p=(2s+1)\lambda/2$\\
\hline
\end{tabular}
\end{center}
\end{table}

So far we have considered the representations for $\lambda>0$ when the lowest weight state has $\mft=s$. Now we take a look at the case when $\lambda<0$ and $\omega$ is then given as
\be
\omega = \half -\lambda(2s+1)
\ee
The action of grade 0 supersymmetry generators on these states produces states with $\mft=s+1/2$. This is different from the case with $\lambda>0$ where we obtained states with $\mft=s-1/2$ by the action of grade 0 supersymmetry generators. Thus the the lowest energy supermultiplets of $SU(2|1)$ that uniquely determine the deformed minimal unitary supermultiplets of $D(2,1;\lambda)$ transform in the representation with the super tableau $\underbrace{\sgenrowbox}_{2s+1}$ which decomposes under the even subgroup 
$SU(2)_{\mathcal{T}} \times U(1)_{\mathcal{J}}$ as
\bea
\underbrace{\sgenrowbox}_{2s+1} = (\underbrace{\genrowbox}_{2s+1}, 0) \oplus (\underbrace{\genrowbox}_{2s}, \frac{\lambda}{2} )
\eea
By acting with the grade +1 supersymmetry generators on these states, we complete the $D(2,1;\lambda)$ supermultiplet given as:
\be
\Psi^{p}_{(s+1/2,0)} \oplus \Psi^{p+1/2}_{(s,1/2)} \oplus \Psi^{p+1}_{(s-1/2,0)} 
\ee
where $p=|\lambda|(2s+1)/2$. We have summarized the deformed supermultiplets for lowest weight states with $\mft=s$ and $\lambda<-1/(4s+2)$ in Table \ref{table-snegative}. 
\begin{table}[htp]
\caption{  \label{table-snegative} Decomposition of $SU(2)$  deformed minimal unitary  lowest energy supermultiplets of $D(2,1;\lambda)$  with respect to $SU(2)_{\mathcal{T}}\times SU(2)_A\times SU(1,1)_K$ . The conformal wavefunctions transforming in the $(\mft,a)$ representation of $SU(2)_{\mathcal{T}}\times SU(2)_A$ with conformal energy $\omega$ are denoted as $\Psi_{(\mft,a)}^{\mathfrak{\omega}}$.
The first column shows the super tableaux  of the lowest energy $SU(2|1)$ supermultiplet, the second column gives the eigenvalue of the $U(1)$ generator $\mathcal{H}$. The allowed range of $\lambda$ in this case is $\lambda<-1/(4s+2)$.} 
\begin{center}
\begin{tabular}{|c|c|c|}
\hline
 $SU(2|1) l.w.v$ & $\mathcal{H}$ & $SU(1,1)_K\times SU(2)_\mathcal{T}\times SU(2)_A$\\
\hline
&& \\
 $\sonebox$ & $|\lambda|/2$ & $\Psi^{|\lambda|/2}_{(1/2,0)} \oplus \Psi^{(|\lambda|+1)/2}_{(0,1/2)} $\\
&& \\
\hline
&& \\
 $\stwobox$ & $|\lambda|$ & $\Psi^{|\lambda|}_{(1,0)} \oplus \Psi^{|\lambda|+1/2}_{(1/2,1/2)} \oplus \Psi^{|\lambda|+1}_{(0,0)}$ \\
&&\\
\hline
&&\\
 $\sthreebox$ & $3|\lambda|/2$ & $\Psi^{3|\lambda|/2}_{(3/2,0)} \oplus \Psi^{(3|\lambda|+1)/2}_{(1,1/2)} \oplus \Psi^{3|\lambda|/2+1}_{(1/2,0)}$ \\
&& \\
\hline
$\vdots$ & $\vdots$ & $\vdots$ \\
$\vdots$ & $\vdots$ & $\vdots$ \\
\hline
&&\\
 $\underbrace{\sgenrowbox}_{2s+1}$ & $(2s+1)|\lambda|/2$ & $\Psi^{p}_{(s+1/2,0)} \oplus \Psi^{p+1/2}_{(s,1/2)} \oplus \Psi^{p+1}_{(s-1/2,0)}$ \\
&& $p=(2s+1)|\lambda|/2$\\
\hline
\end{tabular}
\end{center}
\end{table}
%%%%%%%%%%%%%%%%%%%%%%%%%%%%%%%%%%%%%%%%%%%%%%%%%%%%%%%%%%%%%%%%%%%%%%
\subsubsection[moving]{$ \bf \ket{F}= \left(\begin{array}{c}\ket{\upa,0}_F   \\ \ket{\dowa,0}_F\end{array}\right) $} \label{flwv}

If we choose the doublet of states  $\ket{F}=\ket{\upa,0}_F$ and $\ket{F}=\ket{\dowa,0}_F$ as part of the lowest energy supermultiplet,   the states $\ket{\omega ,\mft,m_\mft,a,m_a}$ satisfying the conditions given in (\ref{grade-1}) will have $\mft=s\pm1/2$ and can be written as
\bea\label{s+1/2}
\ket{-\lambda(2s+1)/2,s+1/2,m_\mft,0,0} \eq \frac{1}{\sqrt{2s+1}}\Big\{\sqrt{s+1/2+m_\mft}\tensor{\upa,0}{s,m_\mft-1/2}{\omega}{0} \nn
&& +\sqrt{s+1/2-m_\mft}\tensor{\dowa,0}{s,m_\mft+1/2}{\omega}{0}\Big\}\nn
\eea
where $\omega = -1/2-\lambda(2s+1)$ with $\lambda<0$. The range of $\lambda$ is determined by the square integrability of the states and the positivity of $\hat{g}^2$. This leads to the following restriction on $\lambda$
\be
\lambda <-\frac{3}{2(2s+1)}
\ee
On the other hand for $\mft=s-1/2$, we have
\bea\label{s-1/2}
\ket{\lambda(2s+1)/2,s-1/2,m_\mft,0,0} \eq \frac{1}{\sqrt{2s+1}}\Big\{\sqrt{s+1/2+m_\mft}\tensor{\dowa,0}{s,m_\mft+1/2}{\omega}{0} \nn
&& -\sqrt{s+1/2-m_\mft}\tensor{\upa,0}{s,m_\mft-1/2}{\omega}{0}\Big\}\nn
\eea
where $\omega = -1/2+\lambda(2s+1)$ with $\lambda>0$. The range of $\lambda$ is determined by the square integrability of the states and the positivity of $\hat{g}^2$. This leads to the following restriction on $\lambda$ 

\be
\lambda >\frac{3}{2(2s+1)}
\ee

Let us now study the simplest case when $s=0$. The lowest energy states that are annihilated by grade -1 generators are $\tensor{\pm1/2,0}{0}{|\lambda|-1/2}{0}$ where $\lambda<0$. The action of grade 0 supersymmetry generators on these states gives:
\bea
\mathfrak{Q}_0 \tensor{\upa,0}{0}{|\lambda|-1/2}{0} &=& \tensor{0,\dowa}{0}{|\lambda|+1/2}{0} \nn
\mathfrak{S}_0 \tensor{\upa,0}{0}{|\lambda|-1/2}{0} &=& 0 \nn
\mathfrak{Q}_0^\dagger \tensor{\upa,0}{0}{|\lambda|-1/2}{0} &=& 0 \nn
\mathfrak{S}_0^\dagger \tensor{\upa,0}{0}{|\lambda|-1/2}{0} &=& 0
\eea

\bea
\mathfrak{Q}_0 \tensor{\dowa,0}{0}{|\lambda|-1/2}{0} &=& 0 \nn
\mathfrak{S}_0 \tensor{\dowa,0}{0}{|\lambda|-1/2}{0} &=& \tensor{0,\dowa}{0}{|\lambda|+1/2}{0}\nn
\mathfrak{Q}_0^\dagger \tensor{\dowa,0}{0}{|\lambda|-1/2}{0} &=& 0 \nn
\mathfrak{S}_0^\dagger \tensor{\dowa,0}{0}{|\lambda|-1/2}{0} &=& 0
\eea
Thus the action of grade 0 supersymmetry generators on  states with $t=1/2$ produce states with $t=0$ , but not  $t=1$ as might be expected. Next we examine the action of grade +1 supersymmetry generators on these states. 
\bea
\mathfrak{Q}_+ \tensor{\upa,0}{0}{|\lambda|-1/2}{0} \eq 0\nn
\mathfrak{S}_+ \tensor{\upa,0}{0}{|\lambda|-1/2}{0} \eq \tensor{0,\upa}{0}{|\lambda|+1/2}{0}
\eea

\bea
\mathfrak{Q}_+ \tensor{\dowa,0}{0}{|\lambda|-1/2}{0} \eq \tensor{0,\upa}{0}{|\lambda|+1/2}{0}\nn
\mathfrak{S}_+ \tensor{\dowa,0}{0}{|\lambda|-1/2}{0} \eq 0
\eea
Thus the complete supermultiplet in this case is
\be
\Psi^{|\lambda|/2}_{(1/2,0)} \oplus \Psi^{(|\lambda|+1)/2}_{(0,1/2)}, \qquad \qquad \lambda<0
\ee
Let us now consider the action of grade 0 and grade+1 supersymmetry generators on states with $s\neq0$ given in (\ref{s+1/2}) and (\ref{s-1/2}). The action of grade 0 generators on $\mft=s+1/2$ states is given as
\bea \label{zero-s+1/2}
\mathfrak{Q}_0\ket{|\lambda|(2s+1)/2,s+1/2,m_\mft,0,0} \eq \sqrt{\frac{s+1/2+m_\mft}{2s+1}}\tensor{0,\dowa}{s,m_\mft-1/2}{\omega+1}{0} \nn
\mathfrak{S}_0 \ket{|\lambda|(2s+1)/2,s+1/2,m_\mft,0,0}\eq \sqrt{\frac{s+1/2-m_\mft}{2s+1}}\tensor{0,\dowa}{s,m_\mft+1/2}{\omega+1}{0} \nn
\mathfrak{Q}^\dagger_0 \ket{|\lambda|(2s+1)/2,s+1/2,m_\mft,0,0} \eq -2\lambda\sqrt{\frac{s+1/2-m_\mft}{2s+1}} (s+1/2+m_\mft) \times\nn
  &&\tensor{0,\upa}{s,m_\mft+1/2}{\omega-1}{0} \nn
\mathfrak{S}^\dagger_0\ket{|\lambda|(2s+1)/2,s+1/2,m_\mft,0,0} \eq -2\lambda\sqrt{\frac{s+1/2+m_\mft}{2s+1}} (s+1/2-m_\mft)\times \nn
&&\tensor{0,\upa}{s,m_\mft-1/2}{\omega-1}{0} 
\eea

Let us now evaluate the action of +1 grade supersymmetry generators on the states with $\mft=s+1/2$.
\bea
\mathfrak{Q}_+\ket{|\lambda|(2s+1)/2,s+1/2,m_\mft,0,0} \eq \sqrt{\frac{s+1/2-m_\mft}{2s+1}}\Big\{\tensor{0,\upa}{s,m_\mft+1/2}{\omega+1}{0}\nn
&& +2\lambda(s+1/2+m_\mft)\tensor{0,\upa}{s,m_\mft+1/2}{\omega-1}{0}\Big\} \nn
\eq \sqrt{\frac{s+1/2-m_\mft}{2s+1}}\Big\{\tensor{0,\upa}{s,m_\mft+1/2}{\omega-1}{1} \nn
&& +(2m_\mft-1)\lambda\tensor{0,\upa}{s,m_\mft+1/2}{\omega-1}{0}
\eea

\bea
\mathfrak{S}_+\ket{|\lambda|(2s+1)/2,s+1/2,m_\mft,0,0} \eq \sqrt{\frac{s+1/2+m_\mft}{2s+1}}\Big\{\tensor{0,\upa}{s,m_\mft-1/2}{\omega+1}{0}\nn
&& +2\lambda(s+1/2-m_\mft)\tensor{0,\upa}{s,m_\mft-1/2}{\omega-1}{0}\Big] \nn
\eq \sqrt{\frac{s+1/2+m_\mft}{2s+1}}\Big\{\tensor{0,\upa}{s,m_\mft-1/2}{\omega-1}{1} \nn
&& -(2m_\mft+1)\lambda\tensor{0,\upa}{s,m_\mft-1/2}{\omega-1}{0}
\eea

\bea
\commute{\mathfrak{Q}_+}{\mathfrak{S}_+}\ket{|\lambda|(2s+1)/2,s+1/2,m_\mft,0,0} \eq 0
\eea
Next we need to evaluate the action of +1 grade supersymmetry generators on the states obtained in (\ref{zero-s+1/2}) which are of the form $\tensor{0,\dowa}{s,m_\mft\pm1/2}{\omega+1}{0}$. From the previous section we would expect states with $\mft=s\pm1/2$ but the states with $\mft=s+1/2$ obtained in this fashion are excitations so the only new states we obtain are the states with $\mft=s-1/2$.

The lowest energy supermultiplet  for $\mft=s+1/2$  corresponds to the following $SU(2|1)$ supertableau
\be
\underbrace{\sgenrowbox}_{2s+1} = \Big(\underbrace{\genrowbox}_{2s+1},1\Big) \oplus \Big(\underbrace{\genrowbox}_{2s},\onebox\Big)
\ee
and the resulting unitary supermultiplet of $D(2,1;\lambda)$ decomposes as 
\be
\Psi^{p}_{(s+1/2,0)} \oplus \Psi^{p+1/2}_{(s,1/2)} \oplus \Psi^{p+1}_{s-1/2,0}
\ee
where $p=(2s+1)|\lambda|/2$ with $\lambda<0$. We have summarized the deformed supermultiplets for lowest weight states with $\mft=s+1/2$ in Table \ref{table-s+1/2}. Note that these occur only for $\lambda <-3/(4s+2)$. 

\begin{table}[htp]
\caption{  \label{table-s+1/2} Decomposition of $SU(2)$  deformed minimal unitary  lowest energy supermultiplets of $D(2,1;\lambda)$  with respect to $SU(2)_{\mathcal{T}}\times SU(2)_A\times SU(1,1)_K$ . The conformal wavefunctions transforming in the $(\mft,a)$ representation of $SU(2)_{\mathcal{T}}\times SU(2)_A$ with conformal energy $\omega$ are denoted as $\Psi_{(\mft,a)}^{\mathfrak{\omega}}$.
The first column shows the super tableaux  of the lowest energy $SU(2|1)$ supermultiplet, the second column gives the eigenvalue of the $U(1)$ generator $\mathcal{H}$. The allowed range of $\lambda$ in this case is $\lambda<-3/(4s+2)$.} 
\begin{center}
\begin{tabular}{|c|c|c|}
\hline
 $SU(2|1) l.w.v$ & $\mathcal{H}$ & $SU(1,1)_K\times SU(2)_\mathcal{T}\times SU(2)_A$\\
\hline
&& \\
 $\sonebox$ & $|\lambda|/2$ & $\Psi^{|\lambda|/2}_{(1/2,0)} \oplus \Psi^{(|\lambda|+1)/2}_{(0,1/2)} $\\
&& \\
\hline
&& \\
 $\stwobox$ & $|\lambda|$ & $\Psi^{|\lambda|}_{(1,0)} \oplus \Psi^{|\lambda|+1/2}_{(1/2,1/2)} \oplus \Psi^{|\lambda|+1}_{(0,0)}$ \\
&&\\
\hline
&&\\
 $\sthreebox$ & $3|\lambda|/2$ & $\Psi^{3|\lambda|/2}_{(3/2,0)} \oplus \Psi^{(3|\lambda|+1)/2}_{(1,1/2)} \oplus \Psi^{3|\lambda|/2+1}_{(1/2,0)}$ \\
&& \\
\hline
$\vdots$ & $\vdots$ & $\vdots$ \\
$\vdots$ & $\vdots$ & $\vdots$ \\
\hline
&&\\
 $\underbrace{\sgenrowbox}_{2s+1}$ & $(2s+1)|\lambda|/2$ & $\Psi^{p}_{(s+1/2,0)} \oplus \Psi^{p+1/2}_{(s,1/2)} \oplus \Psi^{p+1}_{(s-1/2,0)}$ \\
&& $p=(2s+1)|\lambda|/2$\\
\hline
\end{tabular}
\end{center}
\end{table}
%%%%%%%%%%%%%%%%%%%%%%%%%%%%%%%%%%%%%%%%%%%%%%%%%%%%%%%%%%%%%%%%%%%%
Next we look at the states with $\mft=s-1/2$. The action of grade 0 generators on these states is given below:
\bea \label{zero-s-1/2}
\mathfrak{Q}_0\ket{\lambda(2s+1)/2,s-1/2,m_\mft,0,0} \eq -\sqrt{\frac{s+1/2-m_\mft}{2s+1}}\tensor{0,\dowa}{s,m_\mft-1/2}{\omega+1}{0}\nn
\mathfrak{S}_0\ket{\lambda(2s+1)/2,s-1/2,m_\mft,0,0}\eq \sqrt{\frac{s+1/2+m_\mft}{2s+1}}\tensor{0,\dowa}{s,m_\mft+1/2}{\omega+1}{0} \nn
\mathfrak{Q}^\dagger_0 \ket{\lambda(2s+1)/2,s-1/2,m_\mft,0,0} \eq 2\lambda\sqrt{\frac{s+1/2+m_\mft}{2s+1}} (s+1/2-m_\mft)\times \nn
&&\tensor{0,\upa}{s,m_\mft+1/2}{\omega-1}{0}\nn
\mathfrak{S}^\dagger_0 \ket{\lambda(2s+1)/2,s-1/2,m_\mft,0,0}\eq -2\lambda\sqrt{\frac{s+1/2-m_\mft}{2s+1}} (s+1/2+m_\mft)\times \nn
&&\tensor{0,\upa}{s,m_\mft-1/2}{\omega-1}{0}
\eea

The action of +1 grade supersymmetry generators on the states with $\mft=s-1/2$ is given as:
\bea
\mathfrak{Q}_+\ket{\lambda(2s+1)/2,s-1/2,m_\mft,0,0}\eq \sqrt{\frac{s+1/2+m_\mft}{2s+1}}\Big\{\tensor{0,\upa}{s,m_\mft+1/2}{\omega+1}{0}\nn
&& -2\lambda(s+1/2-m_\mft)\tensor{0,\upa}{s,m_\mft+1/2}{\omega-1}{0}\Big\} \nn
\eq \sqrt{\frac{s+1/2+m_\mft}{2s+1}}\Big\{\tensor{0,\upa}{s,m_\mft+1/2}{\omega-1}{1} \nn
&& +(2m_\mft-1)\lambda\tensor{0,\upa}{s,m_\mft+1/2}{\omega-1}{0}
\eea

\bea
\mathfrak{S}_+\ket{\lambda(2s+1)/2,s-1/2,m_\mft,0,0} \eq -\sqrt{\frac{s+1/2-m_\mft}{2s+1}}\Big\{\tensor{0,\upa}{s,m_\mft-1/2}{\omega+1}{0}\nn
&& -2\lambda(s+1/2+m_\mft)\tensor{0,\upa}{s,m_\mft-1/2}{\omega-1}{0}\Big] \nn
\eq \sqrt{\frac{s+1/2-m_\mft}{2s+1}}\Big\{\tensor{0,\upa}{s,m_\mft-1/2}{\omega-1}{1} \nn
&& -(2m_\mft+1)\lambda\tensor{0,\upa}{s,m_\mft-1/2}{\omega-1}{0}
\eea

\bea
\commute{\mathfrak{Q}_+}{\mathfrak{S}_+}\ket{\lambda(2s+1)/2,s-1/2,m_\mft,0,0}\eq 0
\eea
Next we need to evaluate the action of +1 grade supersymmetry generators on the states obtained in (\ref{zero-s-1/2}) which are of the form $\tensor{0,\dowa}{s,m_\mft\pm1/2}{\omega+1}{0}$. From the previous section we would expect states with $\mft=s\pm1/2$ but the states with $\mft=s-1/2$ obtained in this fashion are excitations so the only new states we obtain are the states with $\mft=s+1/2$.

%%%%%%%%%%%%%%%%%%%%%%%%%%%%%%%%%%%%%%%%%%%%%%%%%%%%%%%%%%%%%%%%%%%%%%

The lowest energy super multiplet  for $\mft=s-1/2$ corresponds to the following $SU(2|1)$ supertableau
\be
\underbrace{\sgenrowonebox}_{2s} = \Big(\underbrace{\genrowonebox}_{2s},1\Big) \oplus \Big(\underbrace{\genrowbox}_{2s},\onebox\Big) 
\ee
and leads to the supermultiplet
\be
\Psi^{p}_{(s-1/2,0)} \oplus \Psi^{p+1/2}_{(s,1/2)}  \oplus \Psi^{p+1}_{(s+1/2,0)}
\ee
where $p=(2s+1)\lambda/2$ with $\lambda>0$. We have summarized the deformed supermultiplets for lowest weight states with $\mft=s-1/2$ in Table \ref{table-s-1/2}. Note that these occur only for $s>1/2$ and $\lambda >3/(4s+2)$.

\begin{table}[htp]
\caption{   Decomposition of $SU(2)$  deformed minimal unitary  lowest energy supermultiplets of $D(2,1;\lambda)$  with respect to $SU(2)_{\mathcal{T}}\times SU(2)_A\times SU(1,1)_K$ . The conformal wavefunctions transforming in the $(\mft,a)$ representation of $SU(2)_{\mathcal{T}}\times SU(2)_A$ with conformal energy $\omega$ are denoted as $\Psi_{(\mft,a)}^{\mathfrak{\omega}}$.
The first column shows the super tableaux  of the lowest energy $SU(2|1)$ supermultiplet, the second column gives the eigenvalue of the $U(1)$ generator $\mathcal{H}$. The allowed range of $\lambda$ in this case is $\lambda>3/(4s+2)$.} 
\label{table-s-1/2}
\begin{center}
\begin{tabular}{|c|c|c|}
\hline
 $SU(2|1) l.w.v$ & $\mathcal{H}$ & $SU(1,1)_K\times SU(2)_\mathcal{T}\times SU(2)_A$\\
\hline
&& \\
$\soneonebox$ & $\lambda$ & $\Psi^{\lambda}_{(0,0)} \oplus \Psi^{\lambda+1/2}_{(1/2,1/2)} \oplus \Psi^{\lambda+1}_{(1,0)} $ \\
\hline
$\stwoonebox$ & $3\lambda/2$ & $\Psi^{3\lambda/2}_{(1/2,0)} \oplus \Psi^{(3\lambda+1)/2}_{(1,1/2)} \oplus \Psi^{3\lambda+1}_{(3/2,0)}$ \\
\hline
$\sthreeonebox$ & $2\lambda$ & $\Psi^{2\lambda}_{(1,0)} \oplus \Psi^{2\lambda+1/2}_{(3/2,1/2)} \oplus \Psi^{2\lambda+1}_{(2,0)}$ \\ 
\hline
$\vdots$ & $\vdots$ & $\vdots$ \\
$\vdots$ & $\vdots$ & $\vdots$ \\
\hline
 $\underbrace{\marcshapirowboxs}_{2s}$ & $(2s+1)\lambda/2$ & $\Psi^{p}_{(s-1/2,0)} \oplus \Psi^{p+1/2}_{(s,1/2)} \oplus \Psi^{\lambda+1}_{(s+1/2,)}$ \\
&& $p=(2s+1)\lambda/2$\\
\hline
\end{tabular}
\end{center}
\end{table}

We note the similarities of Table \ref{table-spositive} with Table \ref{table-s-1/2}, and that of Table \ref{table-snegative} with Table \ref{table-s+1/2}. This shows that the supermultiplets obtained for lowest weight states with $\mft=s$ and $\mft=s-1/2$ and $\lambda>0$ are the same and the supermultiplets for lowest weight states with $\mft=s$ and $\mft=s+1/2$ and $\lambda<0$ are same. The difference between these two types of supermultiplets is that the $SU(1,1)_K$ spin (labeled as $p$) gets interchanged between states with $\mft=s+1/2$ and $\mft=s-1/2$ as we change the sign of $\lambda$.
%%%%%%%%%%%%%%%%%%%%%%%%%%%%%%%%%%%%%%%%%%%%%%%%%%%%%%%%%%%%%%%%%%%%%%
\section{SU(2) Deformations of  the minimal unitary representation of $D(2,1;\lambda)$ using both bosons and fermions and $OSp(2n^*|2m)$ superalgebras}

Above we obtained unitary supermultiplets of $D(2,1;\lambda)$ which are $SU(2)$ deformations of the minimal unitary representation. This was achieved by introducing bosonic oscillators $a_n$ and $b_n$ and extending the $SU(2)_T$ generators to  the generators of the diagonal subgroup of $SU(2)_T$ and $SU(2)_S$ realized as bilinears of the bosonic oscillators
\bea S_+ = a^n b_n \\
S_- =b^n a_n \\
S_0 =\frac{1}{2} (N_a - N_b) 
\eea
As stated above, the noncompact group $SO^*(2n)$ generated by the bilinears
\bea
A_{mn} = a_m b_n -a_n b_m \\
A^{mn} = a^m b^n - a^n b^m \\
U^m_n = a^m a_n + b_n b^m 
\eea
commutes with the generators of $D(2,1;\lambda)$. One can similarly obtain $SU(2)$ deformations of the minimal unitary supermultiplet of $D(2,1;\lambda)$ by introducing fermionic oscillators $\rho_r $ and $\sigma_s$ $( r,s,..=1,..n)$ satisfying
\be
\anticommute{\rho_r}{\rho^s}=\anticommute{\sigma_r}{\sigma^s} = \delta^s_r \qquad \anticommute{\rho_r}{\rho_s}=\anticommute{\rho_r}{\sigma_s} \anticommute{\sigma_r}{\sigma_s} =0
\ee
and extend the generators of $SU(2)_T$ to the generators of the diagonal subgroup of $SU(2)_T$ and $SU(2)_F$ generated by
\bea
F_+ = \rho^r \sigma_r \\
F_- = \sigma^r \rho_r \\
F_0 = \frac{1}{2}( \rho^r \rho_r - \sigma^r \sigma_r )
\eea
In this case the  compact $USp(2n)$ generated by the fermion bilinears
\bea
S_{rs} = \rho_{r} \sigma_{s} + \rho_{s} \sigma_{r} \\
S^{rs} = \sigma^{r} \rho^{s} + \sigma^{s} \rho^{r} \\
S^r_s = \rho^r \rho_s - \sigma_s \sigma^r 
\eea
commute with the generators of $D(2,1;\lambda)$. 
One can deform the minimal unitary representation of  $D(2,1;\lambda)$ using fermions and bosons simultaneously. This  is achieved by replacing the $SU(2)_T$ generators by the diagonal generators of $SU(2)_T\times SU(2)_S\times SU(2)_F$ , which we shall denote as $U_+ ,  U_- $ and $U_0$
\bea
U_+ = T_+ + S_+ + F_+ \\
U_- =T_- + S_- + F_- \\
U_0 = T_0 + S_0 + F_0 
\eea
and substituting the quadratic Casimir of $SU(2)_T$ in the Ansatz for $K_-$ with the quadratic Casimir of $SU(2)_D$. Remarkably, in this case the resulting generators of $D(2,1;\lambda)$ commute with the generators of the noncompact superalgebra $OSp(2n^*|2m)$  generated by the  generators of $SO^*(2n)$ and $USp(2m)$ given above and the supersymmetry generators:
\be
\Pi_{mr} = a_m \sigma_r - b_m \rho_r,  \qquad\qquad  \bar{\Pi}^{mr}=(\Pi_{mr})^\dagger = a^m \sigma^r - b^m \rho^r
\ee
\be
\Sigma_m^r = a_m \rho^r + b_m \sigma^r,  \qquad\qquad  \bar{\Sigma}^m_r=(\Sigma_m^r)^\dagger = a^m \rho_r + b^m \sigma^r
\ee
The (anti) commutation relations for the $OSp(2n^*|2m)$ algebra are given below:
\be
\begin{array}{cclccl}
\commute{A^i_j}{A^k_l} \eq \delta^k_j A^i_l -\delta^i_l A^k_j, & \commute{S^r_s}{S^t_u} \eq \delta^t_s S^r_u -\delta^r_u S^t_s \nn
\commute{A_{ij}}{A^k_l} \eq \delta^k_j A_{il} -\delta^k_i A_{jl}, & \commute{S_{rs}}{S^t_u} \eq \delta^t_s S_{ru} +\delta^t_r S_{us} \nn
\commute{A^{ij}}{A^k_l} \eq \delta^i_l A^{jk} -\delta^j_l A^{ik}, & \commute{S^{rs}}{S^t_u} \eq -\delta^s_u S^{rt} -\delta^r_u S^{ts} \nn
\end{array}
\ee
\be\nonumber
\commute{A_{ij}}{A^{kl}} = -\delta^k_j A^l_i + \delta^l_j A^k_i - \delta^l_i A^k_j + \delta^k_i A^l_j
\ee
\be\nonumber
\commute{S_{rs}}{S^{tu}} = -\delta^t_s S^u_r - \delta^t_r S^u_s - \delta^u_s S^t_r - \delta^u_r S^t_s
\ee

\be
\begin{array}{cclccl}
\anticommute{\Pi_{mr}}{\bar{\Pi}^{ns}} \eq \delta^s_r A^n_m- \delta^n_m S^s_r, & \anticommute{\Sigma_m^{r}}{\bar{\Sigma}^n_{s}} \eq \delta^r_s \, A^n_{m} + \delta^n_m \, S^r_{s} \\
\anticommute{\Pi_{mr}}{\Sigma_n^{s}} \eq \delta^s_{r}A_{mn}, & \anticommute{\Pi_{mr}}{\bar{\Sigma}^n_{s}}
\eq - \delta^n_m \, S_{rs} \\
\commute{A^m_n}{\Pi_{kr}} \eq - \delta^m_k \Pi_{nr}, & \commute{A^m_n}{\Sigma_k^r} \eq - \delta^m_k\Sigma_n^r \\
\commute{A^{mn}}{\Pi_{kr}}
\eq -\delta^m_k \bar{\Sigma}^n_r+\delta^n_k \bar{\Sigma}^m_r, & \commute{A^{mn}}{\Sigma_k^r} \eq -\delta^n_k \bar{\Pi}^m_r+\delta^m_k \bar{\Pi}^n_r \\
\commute{A_{mn}}{\Pi_{kr}} \eq 0, & \commute{A_{mn}}{\Sigma_k^r} \eq 0 \\
\commute{S^r_s}{\Pi_{mt}} \eq - \delta^r_t \Pi_{ms}, & \commute{S^r_s}{\Sigma_m^t} \eq \delta^t_s \Sigma_m^r\\
\commute{S^{rs}}{\Pi_{mt}}\eq \delta^s_t \Sigma^r_m + \delta^r_t \Sigma^s_m,&  \commute{S^{rs}}{\Sigma_m^t} \eq 0 \nn
\commute{S_{rs}}{\Pi_{mt}}\eq 0, & \commute{S_{rs}}{\Sigma_m^t} \eq - \delta^t_r \Pi_{ms}
- \delta^t_s \Pi_{mr}  \nn
\end{array}
\ee

%\newpage
\section{$SU(2)$ deformed minimal unitary supermultiplets of $D(2,1;\alpha)$ and $N=4$ superconformal mechanics  }
\numberwithin{equation}{section}
\subsection{$N=4$ Superconformal Quantum Mechanical Models}
A new class of $N=4$ supersymmetric Calogero-type models have been studied  by various authors in recent years \cite{Fedoruk:2011aa,Fedoruk:2009xf} which are invariant under the 
 superconformal algebra  $D(2,1;\alpha)$.

The construction of $D(2,1;\alpha)$ mechanics and quantization was performed in \cite{Fedoruk:2009xf}. In this section we will review that construction following \cite{Fedoruk:2009xf}
 The on shell component action was shown to take the form \cite{Fedoruk:2009xf}
\begin{eqnarray}
S &=& S_b+ S_f\,, \label{4N-ph}\\
S_b &=&  \int dt \,\Big[\dot x\dot x  + {\textstyle\frac{i}{2}}\, \left(\bar z_k \dot z^k -
\dot{\bar z}_k z^k\right)-\frac{\alpha^2(\bar z_k
z^{k})^2}{4x^2} -A \left(\bar z_k z^{k} -c \right) \Big] \,,\label{bose}\\
S_f &=&  -i \!\int\! dt \left( \bar\psi_k \dot\psi^k -\dot{\bar\psi}_k \psi^k \right) +
2\alpha  \!\int\! dt \, \frac{\psi^{i}\bar\psi^{k} z_{(i} \bar z_{k)}}{x^2} +
\textstyle{\frac{2}{3}}\,(1+2\alpha) \!\displaystyle{\int}\! dt\,
\frac{\psi^{i}\bar\psi^{k} \psi_{(i}\bar\psi_{k)}}{x^2}  . \label{fermi}
\end{eqnarray}
Here $x$, $z^i$ and $\psi^j$ ($i,j = 1,2$) are $d=1$ bosonic and fermionic ``fields", respectively. The fields $z^i$ form a complex doublet of the R-symmetry group $SU(2)$. The last term in (\ref{bose}) represents the constraint 
\begin{equation}\label{con}
\bar z_k z^{k}=c\, ,
\end{equation}
and $A$ is the Lagrange multiplier.

Upon quantization of the action given in (\ref{4N-ph}), the dynamical variables were promoted to quantum mechanical operators with following commutators:
\begin{equation}\label{cB}
[X, P] = i\,, \qquad [Z^i, \bar Z_j] = \delta^i_j \,, \qquad \{\Psi^i,
\bar\Psi_j\}=
-{\textstyle\frac{1}{2}}\,\delta^i_j \, \quad (i,j = 1,2).
\end{equation}
As the action is invariant under the group $D(2,1;\alpha)$, the corresponding  symmetry generators  can be obtained by the Noether procedure. The results as given in  \cite{Fedoruk:2009xf} are:
\begin{equation}\label{Q-qu}
\mathbf{Q}^i =P \Psi^i+ 2i\alpha\frac{Z^{(i} \bar Z^{k)}\Psi_k}{X}+
i(1+2\alpha)\frac{\langle\Psi_{k} \Psi^{k}\bar\Psi^i\rangle}{X}\, ,
\end{equation}
\begin{equation}\label{Qb-qu}
\bar{\mathbf{Q}}_i=P \bar\Psi_i- 2i\alpha\frac{Z_{(i} \bar Z_{k)}\bar\Psi^k}{X} +
i(1+2\alpha)\frac{\langle\bar\Psi^{k} \bar\Psi_{k}\Psi_i\rangle}{X}\,,
\end{equation}
\begin{equation}\label{S-qu}
\mathbf{S}^i =-2\,X \Psi^i + t\,\mathbf{Q}^i,\qquad \bar{\mathbf{S}}_i=-2\,X
\bar\Psi_i+ t\,\bar{\mathbf{Q}}_i\,.
\end{equation}
\begin{eqnarray}
\mathbf{H} &=&{\textstyle\frac{1}{4}}\,P^2  +\alpha^2\frac{(\bar Z_k Z^{k})^2+2\bar Z_k
Z^{k}}{4X^2} - 2\alpha    \frac{Z^{(i} \bar Z^{k)} \Psi_{(i}\bar\Psi_{k)}}{X^2} \label{H-qu}\\
&& -\, (1+2\alpha)
\frac{\langle\Psi_{i}\Psi^{i}\,\bar\Psi^{k} \bar\Psi_{k}\rangle}{2X^2}+
  \frac{(1+2\alpha)^2}{16X^2}\,,\nonumber
\\ \label{K-qu}
\mathbf{K} &=&X^2  - t\,{\textstyle\frac{1}{2}}\,\{X, P\} +
    t^2\, \mathbf{H}\,,
\\ \label{D-qu}
\mathbf{D} &=&-{\textstyle\frac{1}{4}}\,\{X, P\} +
    t\, \mathbf{H}\,,
\\ \label{T-qu}
\mathbf{J}^{ik} &=& i\left[ Z^{(i} \bar Z^{k)}+
2\Psi^{(i}\bar\Psi^{k)}\right]\,,
\\ \label{I-qu}
\mathbf{I}^{1^\prime 1^\prime} &=& -i\Psi_k\Psi^k\,,\qquad
\mathbf{I}^{2^\prime
2^\prime} =
i\bar\Psi^k\bar\Psi_k\,,\qquad \mathbf{I}^{1^\prime 2^\prime}
=-{\textstyle\frac{i}{2}}\,
[\Psi_k,\bar\Psi^k]\,.
\end{eqnarray}
where $t$ is time variable and the symbol $\langle...\rangle$ denotes Weyl ordering:
$$
\langle\Psi_{k} \Psi^{k}\bar\Psi^i\rangle =\Psi_{k}
\Psi^{k}\bar\Psi^i+{\textstyle\frac12}\,\Psi^i\,,\qquad \langle\bar\Psi^{k}
\bar\Psi_{k}\Psi_i\rangle= \bar\Psi^{k} \bar\Psi_{k}\Psi_i +{\textstyle\frac12}\,
\bar\Psi_i
$$
$$
\langle\Psi_{i}\Psi^{i}\,\bar\Psi^{k} \bar\Psi_{k}\rangle
={\textstyle\frac12}\,\left\{\Psi_{i}\Psi^{i},\bar\Psi^{k} \bar\Psi_{k}\right\}
-{\textstyle\frac{1}{4}} =\Psi_{i}\Psi^{i}\,\bar\Psi^{k} \bar\Psi_{k} -
\Psi_{i}\bar\Psi^{i}  +{\textstyle\frac{1}{4}} \,,
$$
and $\bar{\mathbf{Q}}_i=-\left(\mathbf{Q}^i\right)^+$,
$\bar{\mathbf{S}}_i=-\left(\mathbf{S}^i\right)^+$.

In the above set of generators $\mathbf{Q}^i$ and $\mathbf{S}^i$ are supertranslation and superconformal boost generators respectively. The generators $\mathbf{H}$, $\mathbf{K}$ and $\mathbf{D}$ are the Hamiltonian, special conformal transformations and dilatation generators respectively and they form an $su(1,1)$ algebra. The remaining generators $\mathbf{J}^{ik}$ and $\mathbf{I}^{i'k'}$ are the generators of two $su(2)$ algebras.  
%%%%%%%%%%%%%%%%%%%%%%%%%%%%%%%%%%%%%%%%%%%%%%%%%%%%%%%%%%%%%%%%%%%%%%
\subsection{Mapping between the harmonic superspace generators of %
$N=4$ superconformal mechanics and generators of deformed minimal unitary representations of $D(2,1;\alpha)$}
A precise correspondence between the Killing potentials of the isometry groups $G$ of $N=2$  sigma models that couple to $4d$ supergravity  in harmonic superspace and the generators of the minimal unitary representations of   $G$ was first shown  in \cite{Gunaydin:2007vc}. It was then suggested that the correspondence could be made concrete and precise by reducing the four dimensional $\mathcal{N}=2$ sigma models to one dimension with eight supercharges and subsequently quantize them to get supersymmetric quantum mechanical models \cite{Gunaydin:2007vc,Gunaydin:2009pk}. The results presented in this paper and the $D(2,1;\alpha)$ superconformal quantum mechanics reviewed in previous section \cite{Fedoruk:2009xf} provides an opportunity to test this proposal.

The basic quantum mechanical operators  in the minimal unitary representation and their deformations was given in section \ref{sect-construction}  are the coordinate $x$ and its momentum $p$ , fermionic oscillators $\alpha^\dagger, \alpha,\beta^\dagger\, \text{and}\, \beta$  and the bosonic oscillators $a^\dagger, a, b^\dagger \,\text{and}\, b$ with the following commutation relations:
\be
\commute{a}{a^\dagger} = 1=\commute{b}{b^\dagger}, \quad \anticommute{\alpha}{\alpha^\dagger} = 1 = \anticommute{\beta}{\beta^\dagger}
\ee
The generators of quantized $N=4$ superconformal mechanics in harmonic superspace go over to the generators of minimal unitary realization of $D(2,1;\lambda) $ deformed by a pair of bosonic oscillators if we make the simple substitutions listed in Table \ref{Mapping}

\begin{table}[htp] 
\caption{ \label{Mapping} Below we give the correspondence between the quantum mechanical operators of $N=4$ superconformal mechanics and the operators of minimal unitary supermultiplet of  $D(2,1;\lambda) $ deformed by a pair of bosons. The $SU(2)$ indices on the left column are raised and lowered by theLevi-Civita tensor $\epsilon_{ij}$ (with $\epsilon_{12}=\epsilon^{21}=1$) .  }
    \begin{tabular}{|c | c|| c |c| } 
        \hline
    Operators of  $ N=4$ Superconformal    &  Operators of minimal unitary representation \\ 
     Mechanics in   Harmonic superspace   &   of $D(2,1;\lambda) $ deformed by a pair of bosons \\ [0.1in] 
     \hline
     & \\
        $\psi^1$       & $-\frac{i}{\sqrt{2}} \alpha^\dagger$    \\[0.1in] 
        $\psi^2$       & $ -\frac{i}{\sqrt{2}} \alpha^\dagger$   \\ [0.1in] 
        $\bar{\psi_1}$ & $ \frac{i}{\sqrt{2}} \alpha$   \\[0.1in] 
        $\bar{\psi_2}$ & $\frac{i}{\sqrt{2}} \beta$    \\[0.1in] 
        $Z^1$          & $- i a ^\dagger$            \\[0.1in] 
        $Z^2$          & $-i b^\dagger$           \\[0.1in]
        $\bar{Z}_1$    & $ i a $            \\[0.1in] 
        $\bar{Z}_2$    & $i b$            \\ [0.1in]
        \hline
    \end{tabular}
\end{table}

 As expected from the results of \cite{Gunaydin:2007vc,Gunaydin:2009pk} we find a one-to-one correspondence between the  symmetry generators of $D(2,1;\alpha)$ superconformal quantum mechanics and the generators of the minimal unitary representations of $D(2,1;\alpha)$ deformed by a pair of bosons, which we present in Table \ref{Generators}. Using this mapping it is easy to see that the quadratic Casimir obtained in equation (4.26) of \cite{Fedoruk:2009xf} is the same as the one obtained by our construction given in (\ref{C2}) for $\mu=4$. The quantum spectra of $N=4$ superconformal mechanics were also studied in \cite{Fedoruk:2009xf} using the realization reviewed in the previous section. To relate   the quantum spectra of these models  to  the minimal unitary realizations of  $D(2,1;\lambda)$ we tabulate the correspondence between the $SU(1,1), \, SU(2)_R, \, SU(2)_L$ quantum numbers of \cite{Fedoruk:2009xf} and  the $SU(1,1)_K, \, SU(2)_\mathcal{T}, \, SU(2)_A$ spins of our construction  in  Table \ref{spectrum}. Using this table  we see that the superfield contents of the quantum spectra of these models as given in Table 2 of \cite{Fedoruk:2009xf} are exactly the same as supermultiplets described in Tables \ref{table-spositive}, \ref{table-snegative}, \ref{table-s+1/2} and \ref{table-s-1/2} above.

%\newpage
\begin{table}[htp]
\caption{\label{Generators} Below we give the mapping between the symmetry generators of $N=4$ superconformal mechanics in harmonic superspace and the minimal unitary representation of $D(2,1;\lambda)$ deformed by a pair of bosons. The first column lists the Symmetry generators of $N=4$ superconformal quantum mechanics in harmonic superspace ($N=4$ SCQM in HSS) and the second column lists the generators for the minimal unitary realization of $D(2,1;\lambda)$ deformed by a pair of bosons.} \begin{center}
    \begin{tabular}{|c | c|}
        \hline
	$N=4$ SCQM in HSS & Deformed Minreps of $D(2,1;\lambda)$ \\ [0.1in]
        \hline
        & \\
        $iI^{1'1'}$ & $A_+$ \\ [0.1in]
        $-iI^{2'2'}$ & $A_-$  \\  [0.1in]
        $iI^{1'2'}$ & $A_0$ \\ [0.1in]
        $-iJ^{11}$   & $\mathcal{T}_+$  \\ [0.1in]
        $iJ^{22}$   & $\mathcal{T}_-$ \\ [0.1in]
        $iJ^{12}$   & $\mathcal{T}_0$ \\ [0.1in]
	$-\frac{i}{\sqrt{2}}Q^1$      & $\widetilde{Q}^1$ \\ [0.15in]
	$-\frac{i}{\sqrt{2}}Q^2$      & $\widetilde{S}^1$ \\ [0.15in]
	$-\frac{i}{\sqrt{2}}\bar{Q}_1$      & $\widetilde{Q}_1$ \\ [0.15in]
	$-\frac{i}{\sqrt{2}}\bar{Q}_2$      & $\widetilde{S}_1$ \\ [0.15in]
	$-\frac{i}{\sqrt{2}}S^1$      & $Q^1$ \\ [0.15in]
	$-\frac{i}{\sqrt{2}}S^2$      & $S^1$ \\ [0.15in]
	$-\frac{i}{\sqrt{2}}S_1$      & $Q_1$ \\ [0.15in]
	$-\frac{i}{\sqrt{2}}S_2$      & $S_1$ \\ [0.15in]
	$2H$ & $K_+$ \\ [0.1in]
	$-2D$ & $\Delta$ \\[0.1in]
	$\frac{1}{2} K $ & $K_-$ \\[0.1in]
	\hline
    \end{tabular}
    \end{center}
\end{table}

\begin{table}[htp]
\caption{\label{spectrum} Below we give the mapping between the quantum numbers of the spectrum of $N=4$ superconformal mechanics as given in \cite{Fedoruk:2009xf} and the minimal unitary supermultiplets  of $D(2,1;\lambda)$ deformed by a pair of bosons.} \begin{center}
    \begin{tabular}{|c | c|}
        \hline
	Quantum spectrum of $N=4$ SCQM & Deformed Minreps of $D(2,1;\lambda)$ \\ [0.1in]
        \hline
        $r_0$ & $p$ \\ [0.1in]
        $j$ & $\mft$  \\  [0.1in]
        $i$ & $a$ \\ [0.1in]
        $c$   & $2s$ \\
	\hline
    \end{tabular}
    \end{center}
\end{table}
\section{Conclusions}
Above we constructed the $SU(2)$ deformed  minimal unitary supermultiplets of $D(2,1;\lambda)$
using quasiconformal methods as was done for the $4d$ and $6d$ superconformal algebras in \cite{Fernando:2009fq,Fernando:2010ia,Fernando:2010dp}. We showed that for deformations obtained by a pair of bosons there exists a precise mapping to the generators and the quantum spectra of the $N=4$ superconformal mechanical models studied recently. This raises the question what kind of superconformal models correspond  to more general deformations of the minimal unitary representations involving an arbitrary numbers of  pairs of bosons and/or pairs of fermions as formulated above. On the basis of the results of \cite{Gunaydin:2007vc,Gunaydin:2009pk} we expect some of these more general deformations to describe the spectra of  supersymmetric quantum mechanical models with  quaternionic K\"ahler target spaces that descend from $4d$, $N=2$ supersymmetric sigma models that couple to $N=2$ supergravity. One would also like to understand the precise connection between the above results and the $d=2$, $N=4$ supersymmetric gauged WZW models  studied in \cite{Gunaydin:1992zg,Gunaydin:1995gv} that extends the results of \cite{Witten:1991mk} on $N=2$ supersymmetric gauged WZW models.
These gauged WZW models correspond to realizations over spaces of the form 
\[ \frac{G_c}{H\times SU(2)} \times SU(2)\times U(1) \]
where $\frac{G_c}{H\times SU(2)}$ is a compact quaternionic symmetric space.
On the other hand the quaternionic K\"ahler manifolds that couple to $4d$, $N=2$ sugra are noncompact.  We hope to address  these problems in a future study.

{\bf Acknowledgements:}  
This work was supported in part by the National
Science Foundation under grants numbered PHY-1213183 and PHY-0855356. Any opinions,
findings and conclusions or recommendations expressed in this
material are those of the authors and do not necessarily reflect the
views of the National Science Foundation.

\newpage
%\bibliography{combined.bib}
\providecommand{\href}[2]{#2}\begingroup\raggedright\endgroup
\end{document}